\documentclass[reprint,twocolumn,showpacs,preprintnumbers,amsmath, aps,pre,amssymb]{revtex4-1}

\usepackage[algoruled]{algorithm2e}

\usepackage{tikz}
\usepackage{graphicx}
\usepackage{caption}
\usepackage{times}
\usepackage{subcaption}

\usetikzlibrary{arrows,shapes,snakes,automata,backgrounds,petri}

\usepackage{expl3}
\ExplSyntaxOn
\newcommand\latinabbrev[1]{
  \peek_meaning:NTF . {
    #1\@}%
  { \peek_catcode:NTF a {
      #1.\@ }%
    {#1.\@}}}
\ExplSyntaxOff

\newcommand\bcmdtab{\noindent\bgroup\tabcolsep=0pt%
  \begin{tabular}{@{}p{10pc}@{}p{20pc}@{}}}
\newcommand\ecmdtab{\end{tabular}\egroup}

 \newcommand{\remove}[1]{}
\usepackage{color}

\newcommand{\nop}[1]{}

\usepackage{color}

\usepackage[nolist]{acronym}

\begin{document}

\label{firstpage}

\title{Thinking Like a Vertex: a Survey of Vertex-Centric Frameworks for Large-Scale Distributed Graph Processing}
\author{Robert Ryan McCune}
 \email{rmccune@nd.edu}
\author{Tim Weninger}
 \email{tweninge@nd.edu}
\author{Greg Madey}
 \email{gmadey@nd.edu}
\affiliation{%
Department of Computer Science and Engineering
Notre Dame, IN 46556
}%
\date{\today}

\begin{abstract}
The vertex-centric programming model is an established computational paradigm recently incorporated into distributed processing frameworks to address challenges in large-scale graph processing.  Billion-node graphs that exceed the memory capacity of standard machines are not well-supported by popular Big Data tools like MapReduce, which are notoriously poor-performing for iterative graph algorithms such as PageRank.  In response, a new type of framework challenges one to ``think like a vertex" (TLAV) and implements user-defined programs from the perspective of a vertex rather than a graph.  Such an approach improves locality, demonstrates linear scalability, and provides a natural way to express and compute many iterative graph algorithms.  These frameworks are simple to program and widely applicable, but, like an operating system, are composed of several intricate, interdependent components, of which a thorough understanding is necessary in order to elicit top performance at scale.  To this end, the first comprehensive survey of TLAV frameworks is presented.  In this survey, the vertex-centric approach to graph processing is overviewed, TLAV frameworks are deconstructed into four main components and respectively analyzed, and TLAV implementations are reviewed and categorized.
\end{abstract}

\maketitle

\section{Introduction}
\label{sec:intro}
The proliferation of mobile devices,  ubiquity of the web, and plethora of sensors has led to an exponential increase in the amount data created, stored, managed, and processed.  In March 2014, an IBM report claimed that 90\% of the world's data had been generated in the last two years \cite{ibmbigdata}.  Big Data characterizes the problems faced by conventional analytics systems with this dramatic expansion of data volume, velocity, and variety.

To address the challenges posed by Big Data, analytical systems are shifting from shared, centralized architectures to distributed, decentralized architectures.  The MapReduce framework, and its open-source variant, Hadoop, exemplifies this effort by introducing a programming model to facilitate efficient, distributed algorithm execution while abstracting away lower-level details \cite{Dean2008}.  Since inception, the Hadoop/MapReduce ecosystem has grown considerably in support of related Big Data tasks.  

However, these distributed frameworks are not suited for all purposes, in many cases can even result in poor performance~\cite{Munagala1999,Cohen2009,Kang2009}. Algorithms that make use of multiple iterations, especially those using graph or matrix data representations, are particularly poorly suited for popular Big Data processing systems. 

Graph computation is notoriously difficult to scale and parallelize, often due to inherent interdependencies within graph data \cite{Lumsdaine2007}.  As Big Data drives graph sizes beyond the memory capacity of a single machine, data must be partitioned to out-of-memory storage or distributed memory.  However, for sequential graph algorithms, which require random access to all graph data, poor locality and the indivisibility of the graph structure cause time- and resource-intensive pointer-chasing between storage mediums in order to access each datum.

In response to these shortcomings, new frameworks based on the {\em vertex-centric programming model} have been developed with the potential to transform the ways in which researchers and practitioners approach and solve certain problems~\cite{Malewicz2010}.  Vertex-centric computing frameworks are platforms that iteratively execute a user-defined program over vertices of a graph.  The user-defined vertex function typically includes data from adjacent vertices or incoming edges as input, and the resultant output is communicated along outgoing edges.  Vertex program kernels are executed iteratively for a certain number of rounds, or until a convergence property is met.  As opposed to the randomly-accessible, ``global" perspective of the data employed by conventional shared-memory sequential graph algorithms, vertex-centric frameworks employ a local, vertex-oriented perspective of computation, encouraging practitioners to ``think like a vertex" (TLAV).

The first published TLAV framework was Google's Pregel system \cite{Malewicz2010}, which, based off of Valiant's Bulk Synchronous Parallel (BSP) model \cite{Valiant1990}, employs synchronous execution.  While not all TLAV frameworks are synchronous, these frameworks are first introduced here within the context of BSP in order to provide foundational understanding of TLAV concepts.

\subsection{Bulk Synchronous Parallel}

After spending a year with Bill McColl at Oxford in 1988, Les Valiant published the seminal paper on the Bulk Synchronous Parallel (BSP) computing model~\cite{Valiant1990} for guiding the design and implementation of parallel algorithms. Initially touted as ``A Bridging Model for Parallel Computation,'' the BSP model was created to simplify the design of software for parallel hardware, thereby ``bridging'' the gap between high-level programming languages and multi-processor systems. 

\begin{figure*}[!ht]
\centering
\includegraphics[width=\textwidth]{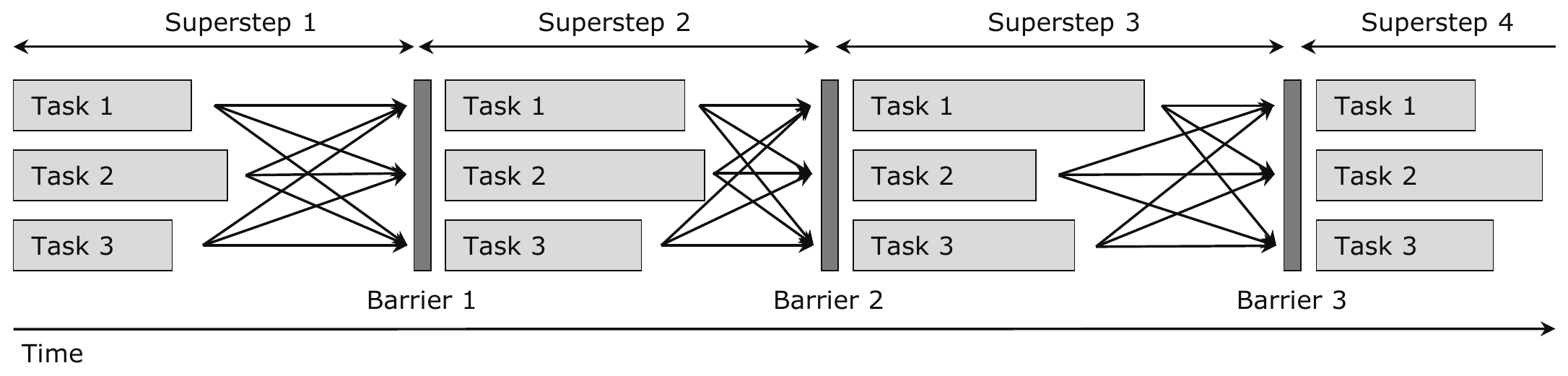}
\caption{Example of Bulk Synchronous Parallel execution with 3 tasks/workers over 4 supersteps. Each task may have varying durations after which messages are passed. The barriers control synchronization across the entire system.}
\label{fig:bsp}
\end{figure*}

As opposed to distributed shared memory or other distributed systems abstractions, BSP makes heavy use of a message passing interface (MPI) which avoids high latency reads, deadlocks and race conditions. BSP is, at the most basic level, a two step process performed iteratively and synchronously: 1) perform task computation on local data, and 2) communicate the results, and then repeat the two steps. In BSP each compute/communicate iteration is called a {\em superstep}, with synchronization of the parallel tasks occurring at the superstep barriers, depicted in Figure~\ref{fig:bsp}.

\subsection{Graph Parallel Systems}

Introduced in 2010, the Pregel system~\cite{Malewicz2010} is a BSP implementation that provides an API specifically tailored for graph algorithms, challenging the programmer to ``think like a vertex.'' Graph algorithms are developed in terms of what each vertex has to compute based on local vertex data, as well as data from incident edges and adjacent vertices.  The Pregel framework, as well other synchronous TLAV implementations, split computation into BSP-style supersteps. Analogous to ``components" in BSP \cite{Valiant1990}, at each superstep a vertex can execute the user-defined vertex function and then send results to neighbors along graph edges. Supersteps always end with a synchronization barrier, shown in Figure~\ref{fig:bsp}, which guarantees that messages sent in a given superstep are received at the beginning of the next superstep. Unlike the original BSP model, vertices may change status between active and inactive, depending on the overall state of execution. Pregel terminates when all vertices halt and no more messages are exchanged.

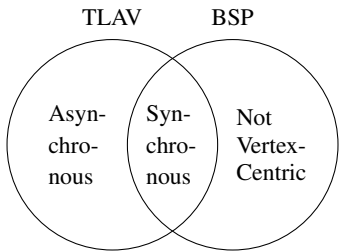
\begin{figure}
\centering
\def\firstcircle{(0,0) circle (1.4cm)}
\def\secondcircle{(0:5cm) circle (1.4cm)}
\def\thirdcircle{(0:1.6cm) circle (1.4cm)}

\begin{tikzpicture}
    \draw \firstcircle node[xshift=0,yshift=17mm] {TLAV};
    \draw node[text width=.5cm,align=center,xshift=-6mm] {Asyn-\\[-1em] chro-\\[-1em] nous};
    \draw node[text width=.5cm,align=center,xshift=7mm] {Syn-\\[-1em] chro-\\[-1em] nous};
    \draw \thirdcircle node [yshift=17mm] {BSP};
    \draw node[text width=.5cm,xshift=19mm] {Not Vertex-Centric};

\end{tikzpicture}

\caption{Comparison of the Think Like a Vertex (TLAV) and Bulk Synchronous Parallel (BSP) models of computation.  Both models are commonly employed for iterative computation.}

\label{fig:bspvstlav}

\end{figure}

A comparison of TLAV frameworks and BSP is presented in Figure~\ref{fig:bspvstlav}.  BSP employs a general model of broad applicability, including graph algorithms at varying levels of granularity.  Underlying BSP execution is the global synchronization barrier among distributed processors.  TLAV frameworks utilize a vertex-centric programming model, and while Pregel and its derivatives employ BSP-founded synchronous execution, other frameworks implement asynchronous execution, which has been demonstrated to improve performance in some instances \cite{Xie2013}.  

In contrast to TLAV and BSP, MapReduce does not natively support iterative algorithms.  Several recent frameworks have extended the MapReduce model to support iterative execution \cite{Kajdanowicz2014}, but for iterative graph algorithms, the graph topological data, which remains static, must be transferred from mappers to reducers, resulting in significant network overhead that renders iterative MapReduce frameworks uncompetitive with TLAV frameworks \cite{Kajdanowicz2014}.  A theoretical comparison between MapReduce and BSP is presented in \cite{bspvsmr}.

\subsection{TLAV Frameworks}

Since Pregel, several TLAV frameworks have been proposed that either employ conceptually alternative framework components (such as asynchronous execution), or improve upon the Pregel model with various optimizations.  This survey provides the first comprehensive examination into TLAV framework concepts, and makes these other contributions:

\begin{enumerate}
\item Analyzes 4 principle components in the design of vertex programs execution in TLAV frameworks, identifying the trade-offs in component implementations and providing data-driven discussion
\item Overviews approaches related to TLAV system architecture, including fault tolerance on distributed systems and novel techniques for large-scale processing on single-machines
\item Discusses how the scalability of a graph algorithm varies inversely with the algorithm's scope, illustrated by vertex-centric and related subgraph-centric, or hybrid, frameworks
\end{enumerate}

This article is organized as follows: First, Section~\ref{sec:overview} overviews the vertex-centric programming model, including an example program and execution.  Section~\ref{sec:sys_theory} presents the four major design decisions, or pillars, of the vertex-centric model. Section~\ref{sec:implementation} presents details for distributed implementation, as well as novel techniques utilized by TLAV frameworks that enable large-scale graph processing on a single machine.  Section~\ref{sec:alt} presents subgraph-centric, or hybrid, frameworks, that adopt a computational scope of the graph that is greater than a vertex (TLAV) but less than the entire graph.  Section~\ref{sec:related} discusses related work.  Finally, Section~\ref{sec:conc} presents a summary, conclusions, and directions for future work.

First, a brief note on terminology:  The TLAV paradigm is described interchangeably as \textit{vertex-centric}, \textit{vertex-oriented}, or \textit{think-like-a-vertex}.  A \textit{vertex program kernel} refers to an instance of the user-defined vertex \textit{program}, \textit{function}, or \textit{process} that is executed on a particular vertex.  A graph is a data structure made up of vertices and edges, both with (potentially empty) data properties.  As in the literature, \textit{graph} and \textit{network} may be used interchangeably, as may \textit{node} and \textit{vertex}, and \textit{edge} and \textit{link}.  \textit{Network} may also refer to hardware connecting two or more machines, depending on context.  A \textit{worker} refers to a slave machine in the conventional master-worker architectural pattern, and a \textit{worker process} is the program that governs worker behavior, including, but not limited to, execution of vertex programs, inter-machine communication, termination, check-pointing, etc.  Graphs are assumed to be directed without loss of generality.

\section{Overview}
\label{sec:overview}

Graph processing is transitioning from centralized to decentralized design patterns.  Sequential, shared-memory graph algorithms are inherently centralized.  Conventional graph algorithms, such as Dijkstra's shortest path \cite{Dijkstra1971} or betweenness centrality \cite{Freeman1977}, receive the entire graph as input, presume all data is randomly accessible in memory ({\em i.e.}, graph-omniscient algorithms), and a centralized computational agent processes the graph in a sequential, top-down manner.  However, the unprecedented size of Big Data-produced graphs, which may contain hundreds of billions of nodes and occupy terabytes of data or more, exceed the memory capacity of standard machines.  Moreover, attempting to centrally compute graph algorithms across distributed memory results in unmanageable pointer-chasing \cite{Lumsdaine2007}.  A more local, decentralized approach is required for processing graphs of scale.

Think like a vertex frameworks are platforms that iteratively execute a user-defined program over vertices of a graph. The vertex program is designed from the perspective of a vertex, receiving as input the vertex's data as well as data from adjacent vertices and incident edges.  The vertex program is executed across vertices of the graph synchronously, or may also be executed asynchronously.  Execution halts after either a specified number of iterations, or all vertices have converged.  The vertex-centric programming model is less expressive than conventional graph-omniscient algorithms, but is easily scalable with more opportunity for parallelism. 

The frameworks are founded in the field of distributed algorithms. Although vertex-centric algorithms are local and bottom-up, they have a provable, global result.  TLAV frameworks are heavily influenced by distributed algorithms theory, including synchronicity and communication mechanisms \cite{Lynch1996}.  Several distributed algorithm implementations, such as distributed Bellman-Ford single-source shortest path \cite{Lynch1996}, are used as benchmarks throughout the TLAV literature.  The recent introduction of TLAV frameworks has also spurred the adaptation of many popular Machine Learning and Data Mining (MLDM) algorithms into graph representations for high-performance TLAV processing of large-scale data sets \cite{Low2010}.

Many graph problems can be solved by both a sequential, shared-memory algorithm as well as a distributed, vertex-centric algorithm.  For example, the PageRank algorithm for calculating web-page importance has a centralized matrix form \cite{Page1999} as well as a distributed, vertex-centric form \cite{Malewicz2010}.  The existence of both forms illustrates that many problems can be solved in more than one way, by more than one approach or computational perspective, and deciding which approach to use depends on the task at hand.  While the sequential, shared-memory approach is often more intuitive and easier to implement on a single machine or centralized architecture, the limits of such an approach are being reached. 

Vertex programs, in contrast, only depend on data local to a vertex, and reduce computational complexity by increasing communication between program kernels.   As a result, TLAV frameworks are highly scalable and inherently parallel, with manageable inter-machine communication.  For example, runtime on the Pregel framework has been shown to scale linearly with the number of vertices on 300 machines \cite{Malewicz2010}.  Furthermore, TLAV frameworks provide a common interface for vertex-program execution, abstracting away low-level details of distributed computation, like MPI, allowing for a fast, re-usable development environment.  A paradigm shift from centralized to decentralized approaches to problem solving is represented by TLAV frameworks.

\subsection{Example: Single Source Shortest Path in TLAV paradigm}
\label{subsec:algo}

The following describes a simple vertex program that calculates the shortest paths from a given vertex to all other vertices in a graph. In contrast to this distributed implementation example, consider a centralized, sequential, shared-memory, or ``graph-omniscient," solution to the single-source shortest path algorithm known as Djikstra's algorithm~\cite{Dijkstra1959} or the more general Bellman–Ford algorithm~\cite{Bellman1958}.

Both Dijkstra's and the Bellman-Ford algorithms are based on repeated relaxations, which iteratively replace distance estimates with more accurate values until eventually reaching the solution. Both variants are have a superlinear time complexity: Djisktra's runs in $O(|E|\log|E|+|V|)$ and Bellman-Ford's runs in $O(|E|\times|V|)$, where $|E|$ is the number of edges and $|V|$ is the number of vertices in the graph and typically $|E|\gg|V|$. Perhaps more importantly, both procedural, shared-memory algorithms keep a large state matrix resulting in a space complexity of $O(|V|^2)$.

\begin{algorithm*}[t]
\small{
\SetKwInOut{Input}{input}
\SetKwFunction{send}{send}\SetKwFunction{receive}{receive}
\SetKwFunction{halt}{halt}\SetKwFunction{min}{min}
\SetAlgoNoLine
    \Input{A graph $(V,E) = G$ with vertices $v\in V$ and edges from $i\rightarrow j$ s.t. $e_{ij}\in E$, \\ and starting point vertex $v_s\in V$}
    \BlankLine
    \lForEach(\tcc*[f]{initialize each vertex data to $\infty$}){$v \in V$}{shrtest\_path\_len$_v\gets \infty$}
    \send(0, $v_s$)\tcc*[r]{to activate, send msg of 0 to starting point}
    \Repeat(\tcc*[f]{The outer loop is synchronized with BSP-styled barriers}){no more messages are sent}{
        \Indp\ForPar(\tcc*[f]{vertices execute in parallel}){$v \in V$}{
            \LinesNumbered
            \Indp\tcc{vertices inactive by default; activated when msg received}
            \tcc{compute minimum value received from incoming neighbors}
            minIncomingData$\gets \min$(\receive(path\_length))\;
            \tcc{set current vertex-data to minimum value}
            \If{\emph{minIncomingData} $<$ \emph{shrtest\_path\_len}$_v$}{
             \Indp   shrtest\_path\_len$_v\gets$ minIncomingData\;
                \ForEach(){$e_{vj} \in E$} {
                \Indp\tcc{send shortest path + edge weight to outgoing edges}
                path\_length $\gets$ shrtest\_path\_len$_v + $weight$_e$\;
                \send(path\_length, $j$)\;}
            }
            \halt()\;
        }
    }
 \caption{Single Source Shortest Path for a Synchronized TLAV Framework}
 \label{alg:sssp}
 }
\end{algorithm*}

In contrast, to solve the same single-source shortest path problem in the TLAV programming model, a vertex program need only pass the minimum value of its incoming edges to its outgoing edges during each superstep. This algorithm, considered a distributed version of Bellman-Ford \cite{Lynch1996}, is shown in Alg.~\ref{alg:sssp}.  The computational complexity of each vertex program kernel is less than that of the sequential solution, however a new dimension is introduced in terms of the communication complexity, or the messaging between vertices \cite{Lynch1996}.  For TLAV implementation, a user need only to write the inner-portion of Alg.~\ref{alg:sssp} denoted by line numbers; the outermost loop and the parallel execution is handled by the framework. Because lines 1-10 are executed on the each vertex these lines are known as the \emph{vertex program}.

The TLAV-solution to the single source shortest path problem has surprisingly few lines of code, and understating its execution requires a different way of thinking.  

\begin{figure*}
\centering

\begin{tikzpicture}[node distance=1.3cm,bend angle=45,auto]
  \tikzstyle{active}=[circle,thick,draw=black!75,fill=white!20,minimum size=6mm]
  \tikzstyle{inactive}=[circle,thick,draw=black!75,fill=gray!20,minimum size=6mm]
  \tikzstyle{every label}=[red]

  \begin{scope}
    \node [inactive] (A1) {$\infty$};
   \node [inactive] (B1) [right of=A1] {$\infty$}
        edge [<->,bend right] node[above] {2} (A1);
	\node [active] (C1) [right of=B1] {0}
	    edge [->,bend left] node[above] {2} (B1);
	\node [inactive] (D1) [right of=C1] {$\infty$}
	    edge [<-, bend right] node[above] {1} (B1)
        edge [<->] node[above] {4} (C1);
    \node (lab1) [right of=D1, xshift=15mm,text width=4cm] {\textit{Superstep 0} \\ \hspace{1mm} message values = 2 and 4};
	
	\node [inactive] (A2) [below of=A1] {$\infty$};
   \node [active] (B2) [right of=A2] {2}
        edge [<->,bend right] (A2)
        edge [<-, dashed] (C1);
	\node [inactive] (C2) [right of=B2] {0}
	    edge [->,bend left] (B2);
	\node [active] (D2) [right of=C2] {4}
	    edge [<-, bend right] (B2)
        edge [<->] (C2)
        edge [<-, dashed] (C1);
    \node (lab2) [right of=D2, xshift=20mm,text width=5cm] {\textit{Superstep 1} \\ \hspace{1mm} message values = 4, 3, and 8};
	
	\node [active] (A3) [below of=A2] {4}
        edge [<-, dashed] (B2);
   \node [inactive] (B3) [right of=A3] {2}
        edge [<->,bend right] (A3);
	\node [inactive] (C3) [right of=B3] {0}
	    edge [<-, dashed] (D2)
	    edge [->,bend left] (B3);
	\node [active] (D3) [right of=C3] {3}
	    edge [<-, bend right] (B3)
        edge [<->] (C3)
        edge [<-, dashed]  (B2);
    \node (lab3) [right of=D3, xshift=15mm,text width=4cm] {\textit{Superstep 2} \\ \hspace{1mm} message values = 6 and 7};
	
	\node [inactive] (A4) [below of=A3] {4} ;
   \node [inactive] (B4) [right of=A4] {2}
        edge [<->,bend right] (A4)
        edge [<-, dashed] (A3);
	\node [inactive] (C4) [right of=B4] {0}
	    edge [<-, dashed] (D3)
	    edge [->,bend left] (B4);
	\node [inactive] (D4) [right of=C4] {3}
	    edge [<-, bend right] (B4)
        edge [<->] (C4); 
    \node (lab4) [right of=D4, xshift=20mm,text width=5cm] {\textit{Superstep 3} \\ \hspace{1mm} Complete, no new messages};
	
  \end{scope}

\end{tikzpicture}

\caption{Computing the Single Source Shortest Path in a graph.  Dashed lines between supersteps represent messages (with values listed to the right), and shaded vertices are inactive.  Edge weights pictorially included in first layer for Superstep 0, then subsequently omitted.} \label{fig:minval}
\end{figure*}
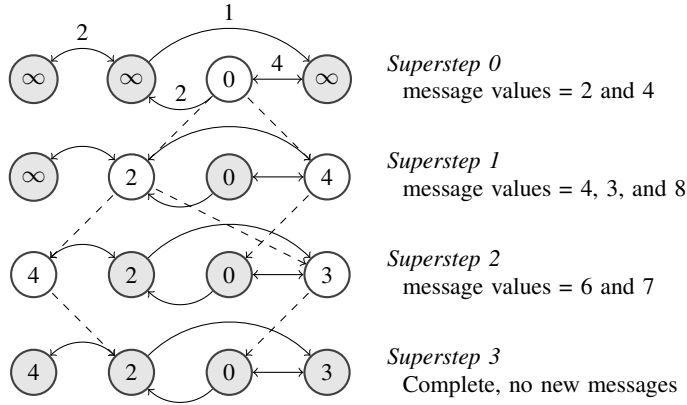

Figure~\ref{fig:minval} depicts the execution of Alg.~\ref{alg:sssp} for a graph with 4 vertices and 6 weighted directed edges. Only the source vertex begins in an active state.  In each superstep, a vertex processes its incoming messages, determines the smallest value among all messages received, and if the smallest received value is less than the vertex's current shortest path, then the vertex adopts the new value as its shortest path, and sends the new path length plus respective edge weights to outgoing neighbors.  If a vertex does not receive any new messages, then the vertex becomes inactive, represented as a shaded vertex in Figure~\ref{fig:minval}.  Overall execution halts once no more messages are sent and all vertices are inactive.

With this example providing insight into TLAV operation, particularly the synchronous message-passing model of Pregel, the survey continues by more completely detailing TLAV properties and categorizing different TLAV frameworks.

\section{Four Pillars of TLAV Frameworks}
\label{sec:sys_theory}

A TLAV framework is software that supports the iterative execution of a user-defined vertex programs over vertices of a graph.  Frameworks are composed of several interdependent components that drive program execution and ultimate system performance.  These frameworks are not unlike an analytic operating system, where component design decisions dictate how computations for a particular topology utilize the underlying hardware.

This section introduces the four principle pillars of TLAV frameworks.  They are:

\begin{enumerate}
\item Timing - How user-defined vertex programs are scheduled for execution
\item Communication - How vertex program data is made accessible to other vertex programs
\item Execution Model - Implementation of vertex program execution and flow of data
\item Partitioning - How vertices of the graph, originally in storage, are divided up to be stored across memory of the system's multiple\footnote{TLAV systems generally distribute a graph across multiple machines because of the graph's prohibitive size.  However, regarding the categorization of ``single machine frameworks" in Section~\ref{sec:arch}, while some TLAV frameworks are implemented for a single machine without the specific intention of developing for a non-distributed environment ( {\em e.g.} the framework is first developed for a single machine before developing the framework for a distributed environment, like the original GraphLab [which published a distributed version 2 years later, see Section~\ref{subsubsec:async}] or GRACE [see Section~\ref{subsubsec:hybrid}]), the single-machine frameworks presented in Section~\ref{sec:arch} are frameworks that implement particularly \textit{novel} methods with the \textit{stated objective} of processing, on a single machine, graphs of size that exceed the single machine's memory capacity.  These single machine frameworks still partition the graph, using framework-specific methods detailed in the respective section.} worker machines
\end{enumerate}

The discussion proceeds as follows: the timing policy of vertex programs is presented in Subsection~\ref{subsec:execution_policy}, where system execution can be synchronous, asynchronous, or hybrid.  Communication between vertex programs is presented in Subsection~\ref{subsec:communication}, where intermediate data is shared primarily through message-passing or shared-memory.  The implementation of vertex program execution is presented in Subsection~\ref{subsec:comp_models}, which overviews popular models of program execution and demonstrates how a particular model implementation impacts execution and performance.  Finally, partitioning of the graph from storage into distributed memory is presented in Subsection~\ref{subsec:partitioning}.

Each pillar is heavily interdependent with other pillars, as each design decision is tightly integrated and strongly influenced by other design decisions.  While each pillar may be understood through a sequential reading of the information provided, a more efficient, yet thorough understanding may be achieved by freely forward- and cross-referencing other pillars, especially when related sections are cited.  The inter-relation of the four pillars is unavoidable and indivisible, \textit{not unlike a graph data structure itself.}  The difficulty of independently describing each pillar certainly reflects the challenge of processing a vertex in which a given result depends on the concurrent processing of neighboring vertices.  This survey is restricted to a sequential presentation of information in the form of a paper.  However, each pillar, though unique, depends on, and may only be described in relation to, other pillars, so a sufficient understanding of any given pillar may only be achieved by understanding all pillars of a TLAV framework, collectively.  Thus one may begin to understand the challenges of processing graphs (especially large graphs, when not all ``pillars" are in the same ``paper") as in Section~\ref{sec:intro}, Section~\ref{sec:overview}, and \cite{Lumsdaine2007}.

\subsection{Timing}
\label{subsec:execution_policy}
In TLAV frameworks, the scheduling and timing of the execution is separate from the logic of the vertex program.  The \textit{timing} of a framework characterizes how active vertices are ordered by the scheduler for computation.  Timing can be synchronous, asynchronous, or a hybrid of the two models.  Frameworks that represent the different fundamental timing models are presented in Table~\ref{table:timing}.

\subsubsection{Synchronous}
\label{subsubsec:sync}
The \textit{synchronous} timing model is based on the original bulk synchronous parallel (BSP) processing model discussed above. In this model, active vertices are executed conceptually in parallel over one or more iterations, called \textit{supersteps}. Synchronization is achieved through a global synchronization \textit{barrier} situated between each superstep that blocks vertices from computing the next superstep until all workers complete the current superstep. Each worker coordinates with the master to progress to the next superstep. Synchronization is achieved because the barrier ensures that each vertex within a superstep has access to only the data from the previous superstep. Within a single processing unit, vertices can be scheduled in a fixed or random order because the execution order does not affect the state of the program.  The global synchronization barrier introduces several performance trade-offs.

Synchronous systems are conceptually simple, demonstrate scalability, and perform exceptionally well for certain classes of algorithms.  While not all TLAV programs consistently converge to the same values depending on system implementation, synchronous systems are almost always deterministic, making synchronous applications easy to design, program, test, debug, and deploy.  Although coordinating synchronization imposes consistent overhead, the overhead becomes largely amortized for large graphs.  Synchronous systems demonstrate good scalability, with runtime often linearly increasing with the number of vertices \cite{Malewicz2010}.  As will be discussed in Section~\ref{subsubsec:msg}, synchronous systems are often implemented along with message-passing communication, which enables a more efficient ``batch messaging'' method. Batch messaging can especially benefit systems with lots of network traffic induced by algorithms with a low computation-to-communication ratio \cite{Xie2013}.

Although synchronous systems are conceptually straight-forward and scale well, the model is not without drawbacks.  One study found that synchronization, for an instance of finding the shortest path in a highly-partitioned graph, accounted for over 80\% of the total running time \cite{Chen}, so system throughput must remain high to justify the cost of synchronization, since such coordination can be relatively costly.  However, when the number of active vertices drops or the workload amongst workers becomes imbalanced, system resources can become under-utilized.   Iterative algorithms often suffer from ``the curse of the last reducer'' otherwise known as the ``straggler'' problem where many computations finish quickly, but a small fraction of computations take a disproportionately longer amount of time \cite{Suri2011}. {\em For synchronous systems, each superstep takes as long as the slowest vertex}, so synchronous systems generally favor lightweight computations with small variability in runtime.

Finally, synchronous algorithms may not converge in some instances.  In graph coloring algorithms, for example, vertices attempt to choose colors different than adjacent neighbors \cite{Gonzalez2011} and require coordination between neighboring vertices. However, during synchronous execution, the circumstance may arise where two neighboring vertices continually flip between each others' color.  In general, algorithms that require some type of neighbor coordination may not always converge with the synchronous timing model without the use of some extra logic in the vertex program \cite{Xie2013}.

\begin{table}[t]
\centering
\small{
\begin{tabular} {l | c l}
Framework & Timing & \\ \hline
Pregel      &   Synchronous & \cite{Malewicz2010} \\
Giraph      & Synchronous   & \cite{Avery2011} \\
Hama        & Synchronous   & \cite{Seo2010} \\
GraphLab    & Asynchronous  & \cite{Low2012,Low2010} \\
PowerGraph  & Both          & \cite{Gonzalez2012} \\ 
PowerSwitch & Hybrid        & \cite{Xie2013} \\
GRACE       & Hybrid        & \cite{Wang2013} \\
GraphHP     & Hybrid        & \cite{Chen} \\
P++         & Hybrid        & \cite{Zhou2014}
\end{tabular}
}
\caption{Execution timing model of selected frameworks.}
\label{table:timing}
\end{table}

\subsubsection{Asynchronous}
\label{subsubsec:async}
In the asynchronous iteration model, no explicit synchronization points, {\em i.e.}, barriers, are provided, so any active vertex is eligible for computation whenever processor and network resources are available.  Vertex execution order can be dynamically generated and reorganized by the scheduler, and the ``straggler'' problem is eliminated. As a result, many asynchronous  models outperform corresponding synchronous models, but at the expense of added complexity.

Theoretical and empirical research has demonstrated that asynchronous execution can generally outperform synchronous execution \cite{Bertsekas1989,Low2012}, albeit precise comparisons for TLAV frameworks depend on a number of properties \cite{Xie2013}.  Asynchronous systems especially outperform synchronous systems when the workload is imbalanced. For example, when computation per vertex varies widely, synchronous systems must wait for the slowest computation to complete, while asynchronous systems can continue execution maintaining high throughput. One disadvantage, however, is that asynchronous execution cannot take advantage of batch messaging optimizations (see Section~\ref{subsubsec:opts}). Thus, synchronous execution generally accommodates I/O-bound algorithms, while asynchronous execution well-serves CPU-bound algorithms by adapting to large and variable workloads.

Many iterative algorithms exhibit asymmetric convergence.  Low {\em et al}. demonstrated that, for PageRank, the majority of vertices converged within one superstep, while only 3\% of vertices required more than 10 supersteps \cite{Low2012}.  Asynchronous systems can utilize prioritized computation via a dynamic schedule to focus on more challenging computations early in execution to achieve better performance \cite{Zhang2011,Low2012}.  Generally, asynchronous systems perform well by providing more execution flexibility, and by adapting to dynamic or variant workloads.

Although intelligent scheduling can improve performance, schedules resulting in sub-optimal performance are also possible.  In some instances, a vertex may perform more updates than necessary to reach convergence, resulting in excessive computation \cite{Zhang2013}.  Moreover, if implementing the pull model of execution, which is commonly implemented in asynchronous systems \cite{Low2012} and described in Section~\ref{subsubsec:pushpull}, communication becomes redundant when neighboring vertex values don't change \cite{Zhang2013,Hant2014}.

The flexibility provided by asynchronous execution comes at the expense of added complexity, not only from scheduling logic, but also from maintaining data consistency.  Asynchronous systems typically implement shared memory, discussed in Section~\ref{subsubsec:shared}, where data race conditions can occur when parallel computations simultaneously attempt to modify the same data.  Additional mechanisms are necessary to ensure mutual exclusion, which can challenge algorithm development because framework users may have to consider low-level concurrency issues \cite{Wang2013}, like, for example, in GraphLab where users must select a consistency model \cite{Low2012}.

\subsubsection{Hybrid}
\label{subsubsec:hybrid}
Rather than adhering to the inherent strengths and weaknesses of a strict execution model, several frameworks work around a particular shortcoming through design improvements.  One such implementation, GraphHP, reduces the high fixed cost of the global synchronization barrier using {\em pseudo-supersteps} \cite{Chen}. Another implementation, GRACE, explores dynamic scheduling within a single superstep \cite{Wang2013}. The PowerSwitch system removes the need to choose between synchronous and asynchronous execution and instead adaptively switches between the two modes to improve performance \cite{Xie2013}.  Together, these three frameworks illustrate how weaknesses with a particular execution model can be overcome through engineering and problem solving, rather than strict adoption of an execution model.

As previously discussed, synchronous systems suffer from the high, fixed cost of the global synchronization barrier.  The hybrid execution model introduced by GraphHP, and also used by P++ framework \cite{Zhou2014}, reduces the number of supersteps by decoupling intra-processor computation from the inter-processor communication and synchronization \cite{Chen}.  To do this GraphHP distinguishes between two types of nodes: \textit{boundary nodes} that share an edge across partitions, and \textit{local nodes} that only have neighboring nodes within the local partition.  During synchronization, messages are only exchanged between boundary nodes. As a result, in GraphHP, a given superstep is composed of two phases: global and local. The global phase, which is executed first, runs the user program across all boundary vertices using data transmitted from other boundary vertices as well as its own local vertices. Once the global phase is complete, the local phase executes the vertex program on local vertices within a pseudo-superstep; the pseudo-superstep is different from a regular superstep in that: 1) pseudo-supersteps have local barriers resulting in local iterations independent of any global synchronization or communication; and 2) local message passing is done through direct, in-memory message passing, which is much faster than standard MPI-style messages.

A similar approach to segmented execution, as in GraphHP and P++, is the KLA paradigm \cite{Harshvardhan2014}, which creates a hybrid of synchronous and asynchronous execution.  For graphs, the depth of asynchronous execution is parameterized, and asynchronous execution is allowed for a certain number of levels before a synchronous round.  Similar to how GraphHP implements a round of boundary vertex execution before several rounds of local execution, KLA has multiple traversals of asynchronous execution before coordinating a round of synchronous execution.  The trade-off is between expensive global synchronizations with cheap but possibly redundant asynchronous computations.  KLA is also similar to delta-stepping used for single source shortest path \cite{Meyer2003}.

The single-machine framework GRACE explores dynamic scheduling of vertices from within a single synchronous round\cite{Wang2013}. To do this GRACE exposes a programming interface that, from within a given superstep, allows for prioritized execution of vertices and selective receiving of messages outside of the previous superstep.  Results demonstrate comparable runtime to asynchronous models, with better scaling across multiple worker threads on a single machine.

Knowing \textit{a priori} which execution mode will perform better for a given problem, algorithm, system, or circumstance is challenging.  Furthermore, the underlying properties that give one execution model an advantage over another may change over the course of processing.  For example, in the distributed Single Source Shortest Path algorithm \cite{Bertsekas1996}, the process begins with few active vertices, where asynchronous execution is advantageous, then propagates to a high number of active vertices performing lightweight computations, which is ideal for synchronous execution, before finally converging amongst few active vertices \cite{Xie2013}.  For some algorithms, one execution mode may outperform another only for certain stages of processing, and the best mode at each stage can be difficult to predict.

Motivated by the necessity for execution mode dynamism, PowerSwitch was developed to adaptively switch between synchronous and asynchronous execution modes \cite{Xie2013}.  Developed on top of the PowerGraph platform, PowerSwitch can quickly and efficiently switch between synchronous and asynchronous execution.  PowerSwitch incorporates throughput heuristics with online sampling to predict which execution mode will perform better for the current period of computation.  Results demonstrate that the PowerSwitch's heuristics can accurately predict throughput, the switching between the two execution modes is well-timed, and overall runtime is improved for a variety of algorithms and system configurations \cite{Xie2013}.

\subsection{Communication}
\label{subsec:communication}

Communication in TLAV frameworks entails how data is shared between vertex programs.  The two conventional models for communication in distributed systems, as well as distributed algorithms, are message passing and shared memory \cite{Yan2013,Lu1995,Lynch1996}.  In message passing systems, data is exchanged between processes through messages, whereas in shared memory systems data for one process is directly and immediately accessible by another process.  This section compares and contrasts message passing and shared memory for TLAV frameworks. A third method of communication, active messages, is also presented. Finally, techniques to optimize distributed message passing are discussed.

\begin{figure*}
    \centering
    \begin{subfigure}[b]{.28\textwidth}
        \centering
        \raisebox{12mm}{\includegraphics[width=\textwidth]{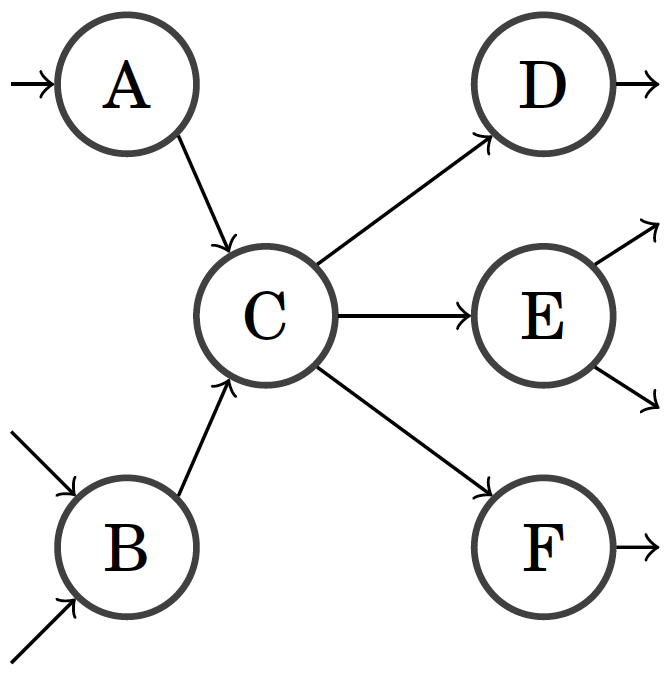}}
        \caption{Sample Graph}
        \label{fig:samp_graph}
    \end{subfigure}
    ~
    \begin{subfigure}[b]{.7\textwidth}
        \centering    
    
        \begin{subfigure}[b]{.43\textwidth}
            \centering
            \includegraphics[width=.75\textwidth]{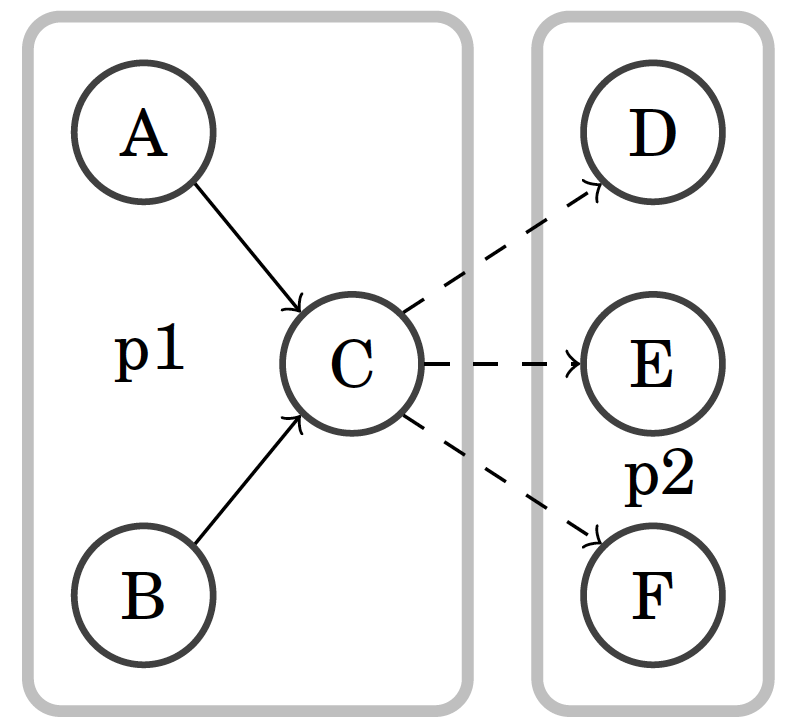}
            \caption{Message Passing}
            \label{fig:pregel}
        \end{subfigure}
        ~
        \begin{subfigure}[b]{.5\textwidth}
            \centering
            \includegraphics[width=\textwidth]{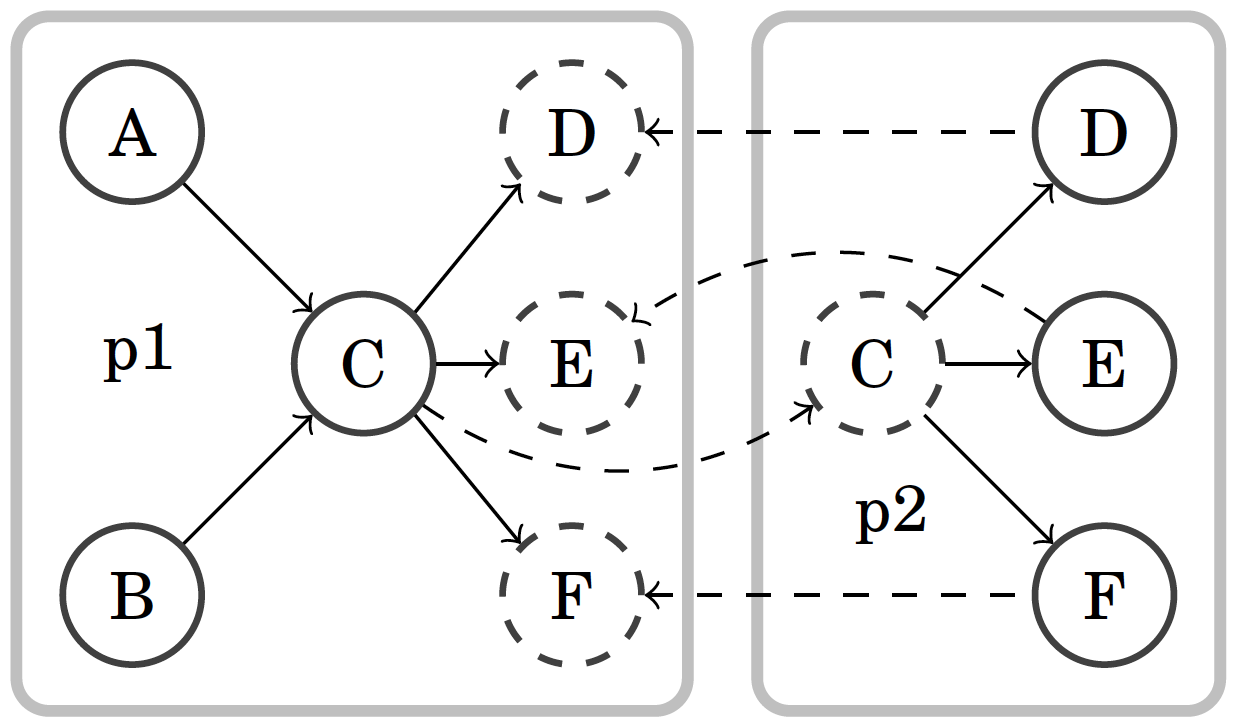}
            \caption{Shared Memory}
            \label{fig:graphlab}
        \end{subfigure}
    
        \begin{subfigure}[b]{.45\textwidth}
            \centering
            \includegraphics[width=.77\textwidth]{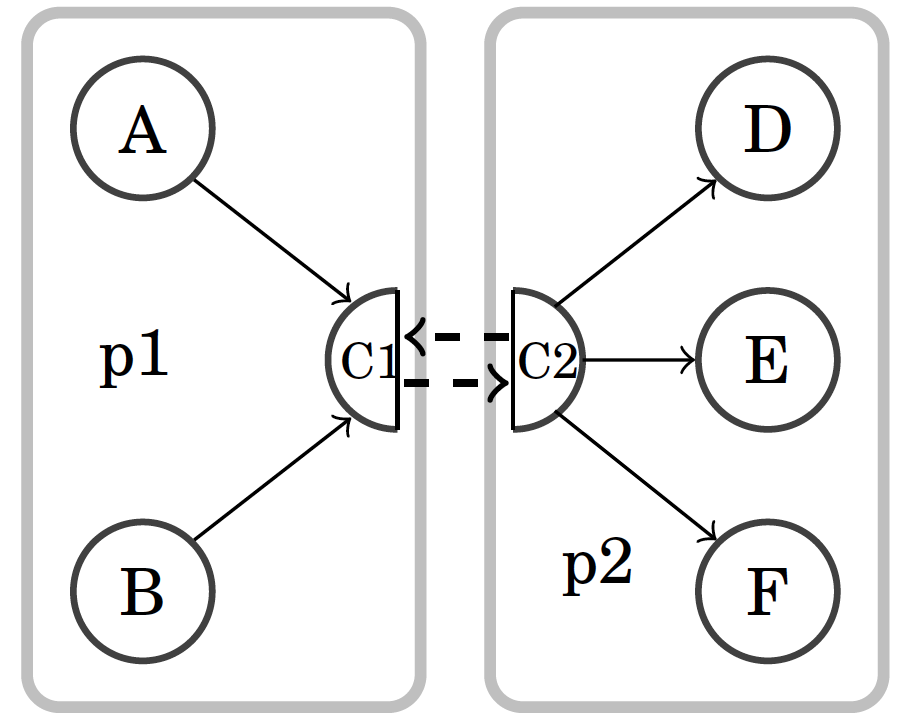}
            \caption{Shared Mem w/Vertex-Cuts}
            \label{fig:powergraph}
        \end{subfigure}
        ~
        \begin{subfigure}[b]{.4\textwidth}
            \centering
            \includegraphics[width=.9\textwidth]{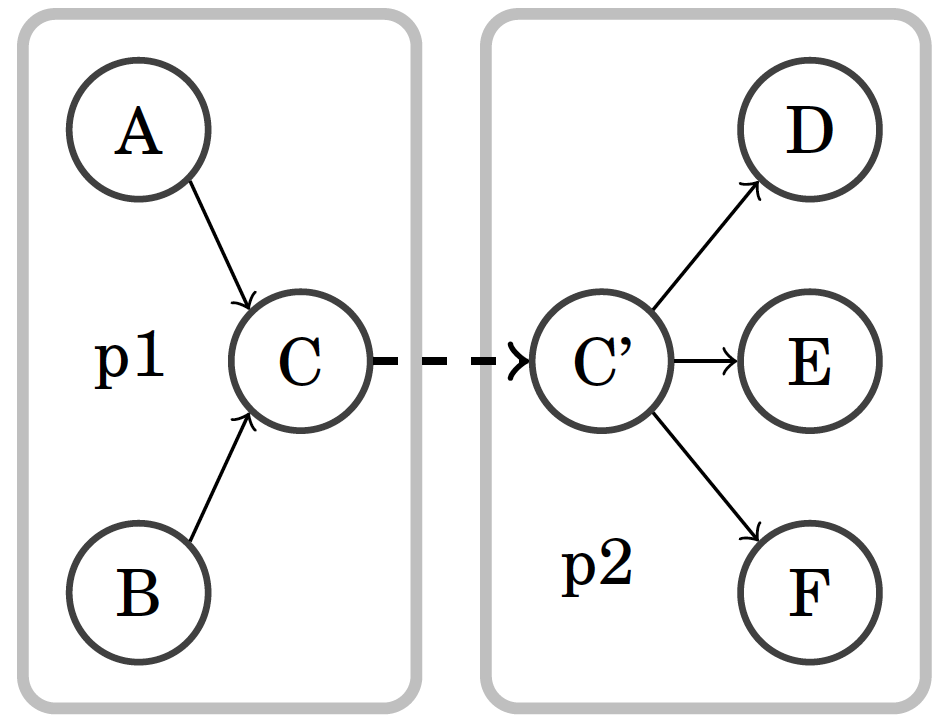}
            \caption{Active Msgs w/Agent-Graph Scatter Vertex}
            \label{fig:gre}
        \end{subfigure}
    
    \end{subfigure}
    
    \caption{Distributed communication patterns for common communication implementations.  The sample graph is partitioned across two machines (see Section~\ref{subsec:partitioning}), with vertices A, B, and C residing on machine p1, and vertices D, E, and F on machine p2.  Pregel is represented in (b), GraphLab in (c), PowerGraph in (d), and GRE in (e).}
    \label{fig:comm_imp}
\end{figure*}

Diagrams in Figure~\ref{fig:comm_imp} are referenced throughout this section to illustrate the different communication implementations.  A sample graph is presented in Figure~\ref{fig:samp_graph}, and Figures~\ref{fig:pregel}-\ref{fig:gre} depict 4 TLAV communication implementations of the sample graph.  For each implementation, vertices are partitioned across 2 machines, namely, vertices A, B, and C are partitioned to machine p1, and vertices D, E, and F are put on machine p2 (except Figure~\ref{fig:powergraph} and \ref{fig:gre}, where the graph is cut along vertex C).  Solid arrows represent local communication\footnote{Local communication means communication between vertices residing on the same machine} and dashed arrows represent network traffic.

\subsubsection{Message Passing}
\label{subsubsec:msg}

In the message passing method of communication, also known as the LOCAL model of distributed computation \cite{Peleg}, information is sent from one vertex program kernel to another via a message. A message contains local vertex data and is addressed to the ID of the recipient vertex.  In the archetypal message-passing framework Pregel \cite{Malewicz2010}, a message can be addressed anywhere, but because vertices do not have ID information of all of other vertices, destination vertex IDs are typically obtained by iterating over outgoing edges.

After computation is complete and a destination ID for each message is determined, the vertex dispatches messages to the local worker process.  The worker process determines whether the recipient resides on the local machine or a remote machine.  In the case of the former, the worker process can place the message directly into the vertex's incoming message queue.  Else, the worker process looks up the worker-id of the destination vertex\footnote{If the graph is partitioned using random hash partitioning, then the destination worker can be determined by hashing the destination vertex ID.  Otherwise, for more advanced partitioning methods (see Section~\ref{subsec:partitioning}), the worker process typically has access to a local routing table, provided by the master during initialization.} and places the message in an outgoing message buffer.  The outgoing message buffer in Pregel, a synchronously-timed system, is flushed when it reaches a certain capacity, sending messages over the network in batches. Waiting until the end of a superstep to send all outgoing remote messages can exceed memory limits \cite{Satish}.

Message passing is commonly implemented with synchronized execution, which guarantees data consistency without low-level implementation details.  All messages sent during superstep $S$ are received in superstep $S+1$, at which point a vertex program can access the incoming message queue at the beginning of $S+1$'s program execution.  Synchronous execution also facilitates batch messaging, which improves network throughput.  For I/O bound algorithms with lightweight computation, such as PageRank \cite{Brin1998}, where vertices are ``always active" so messaging is high \cite{Shang2013}, synchronous execution has been shown to significantly outperform asynchronous execution \cite{Xie2013}.

Message passing is depicted in Figure~\ref{fig:pregel}, where vertex $C$ sends (an) inter-machine message(s) to vertices $D$, $E$, and $F$.  Technically, messages are first sent from $C$ to the worker process of $p1$, which routes the messages to worker process $p2$, which places the message in a vertex's incoming message queue, but the worker process-related routing is omitted from the figure without loss of generality.  Figure~\ref{fig:pregel} represents a general message passing framework, such as Pregel or Giraph.  The three messages sent by $C$ across the network can be potentially reduced using optimization techniques in Section~\ref{subsubsec:opts}, namely, Receiver-side Scatter, depicted in Figure~\ref{fig:rec_scat}.  

\subsubsection{Shared Memory}
\label{subsubsec:shared}

Shared memory exposes vertex data as shared variables that can be directly read or be modified by other vertex programs.  Shared memory avoids the additional memory overhead constituted by messages, and doesn't require intermediate processing by workers.  Shared memory is often implemented by TLAV frameworks developed for a single machine (see Section~\ref{sec:arch}), since challenges to a shared memory implementation arise in the distributed setting \cite{Protic1998,Nitzberg1991}, where consistency must be guaranteed for remotely-accessed vertices.  Inter-machine communication for distributed shared memory still occurs through network messages.  The Trinity framework \cite{Shao2013} implements a shared global address space that abstracts away distributed memory.

For shared memory TLAV frameworks, race conditions may arise when an adjacent vertex resides on a remote machine.  Shared memory TLAV frameworks often ensure memory consistency through mutual exclusion by requiring serializable schedules. Serializability, in this case, means that every parallel execution has a corresponding sequential execution that maintains consistency, {\em cf.}, the dining philosophers problem \cite{Low2012,Gonzalez2012}.  

In GraphLab \cite{Low2012} border vertices are provided locally-cached \textit{ghost} copies of remote neighbors, where consistency between ghosts and the original vertex is maintained using pipelined distributed locking \cite{Dijkstra1971}.  In PowerGraph \cite{Gonzalez2012}, the second generation of GraphLab, graphs are partitioned by edges and cut along vertices (see vertex-cuts in Section~\ref{subsec:partitioning}), where consistency across cached \textit{mirrors} of the cut vertex is maintained using parallel Chandy-Misra locking \cite{Chandy1984}.  GiraphX is a Giraph derivative with a synchronous shared memory implementation \cite{Tasci2013}, which again provides serialization through Chandy-Misra locking of border vertices, although without local cached copies.  The reduced overhead of shared memory compared to message passing is demonstrated by GiraphX, which converges 35\% faster than Giraph when computing PageRank on a large Web Graph \cite{Tasci2013}.  Moreover, some iterative algorithms perform better under serialized conditions, such as Dynamic ALS \cite{Zhou2008,Low2012}, and popular Gibbs sampling algorithms that actually require serializability for correctness \cite{Gonzalez2011}.  

Shared memory implementations are depicted in Figure~\ref{fig:graphlab} and Figure~\ref{fig:powergraph}.  In Figure~\ref{fig:graphlab}, ghost vertices, represented by dashed circles, are created for every neighboring vertex residing on a remote machine, as implemented by GraphLab \cite{Low2012}.  One disadvantage of shared-memory frameworks is seen when computing on scale-free graphs which have a certain percentage of high degree vertices, such as vertex $C$. In these cases the graph can be difficult to partition \cite{Leskovec2009} resulting in many ghost vertices.

Figure~\ref{fig:powergraph} depicts shared memory with vertex cuts as implemented by PowerGraph \cite{Gonzalez2012}.  PowerGraph combines vertex-cuts (discussed in Section~\ref{subsec:partitioning}) with the three-phase Gather-Apply-Scatter computational model (see Section~\ref{subsubsec:gas}) to improve processing of scale-free graphs.  In Figure~\ref{fig:powergraph}, the graph is cut along vertex $C$, where $C1$ is arbitrarily chosen as the master and $C2$ as the mirror.  For each iteration, a distributed vertex preforms computation where: (i) both $C1$ and $C2$ compute a partial result based on local neighbors, (ii) the partial result is sent over the network from the mirror $C2$ to the master $C1$, (iii) the master computes the final result for the iteration, (iv) the master transmits the result back to the mirror over the network, then (v) the result is sent to local neighbors as necessary.  PowerGraph demonstrates how the combination of advanced components, {\em i.e.}, vertex-cuts and three-phase computation, can overcome processing challenges like imbalances arising from high-degree vertices in scale-free graphs.  

Shared memory systems are often implemented with asynchronous execution.  Although consistency is fundamentally maintained in synchronous message passing frameworks like Pregel, asynchronous, shared memory frameworks like GraphLab may execute faster because of prioritized execution and low communication overhead, but at the expense of added complexity for scheduling and maintaining consistency.  The added complexity challenges scalability, for as the number of machines and partitions increase, more time and resources become devoted to locking protocols.  

Dynamic computation addresses asymmetric convergence by only updating necessary vertices.  Shared memory with asynchronous execution is an effective platform for dynamic computation, because the movement of data is separated from computation, allowing vertices to access neighboring values even if the values haven't changed between iterations.  This implies the {\em pull} mode of information flow \ref{subsubsec:pushpull}.  In contrast, a vertex in a message-passing framework would need all neighboring values delivered in order to perform an update, even if some values had not changed.  Dynamic computation is possible with message passing in the Cyclops framework, which implements a {\em distributed immutable view}.  Cyclops is a synchronous shared memory framework \cite{Chen2014}, where one of the replicated vertices is designated the master, which computes updates and messages the updated state to replicas at the end of an iteration.  Cyclops outperforms synchronous message passing frameworks by reducing the amount of processing performed by each worker parsing messages, and is comparable to PowerGraph by delivering significantly fewer messages.

Significant deterioration in performance was noted in \cite{Han2014,Lu2014} for larger graphs, although admittedly performance largely depends on algorithm behavior \cite{Xie2013,Shang2013}.  In short, asynchronous shared memory systems can potentially outperform synchronous message passing systems, though the latter often demonstrate better scalability and generalization.

\subsubsection{Active Messages}
\label{subsubsec:acivemsg}

While message passing and shared memory are the two most commonly implemented forms of communication in distributed systems, a third method called {\em active messages} is implemented in the GRE framework \cite{Yan2013}\footnote{Active messages as described in this section are different from message-passing messages that activate a vertex, like in Pregel}.  Active messaging is a way of bringing computation to data, where a message contains both data as well as the operator to be applied to the data \cite{VonEicken1992}.  Active messages are sent asynchronously, and executed upon receipt by the destination vertex.  Within the GRE architecture, active messages combine the process of sending and receiving messages, removing the need to store intermediate state, like message queues or edge data.  When combined with the framework's novel Agent-Graph model, described below, GRE demonstrates 20\%--55\% reduction in runtime compared to PowerGraph across three benchmark algorithms in real and synthetic datasets, including 39\% reduction in the execution time per iteration for PageRank on the Twitter graph when scaled across 192 cores over 16 machines when compared to a PowerGraph implementation on 512 cores across 64 machines \cite{Yan2013}.

The GRE framework modifies the data graph into an Agent-Graph.  The Agent-Graph is a model used internally by the framework, but is not accessible to the user.  The Agent-Graph adds \textit{combiner} and \textit{scatter} vertices to the original graph in order to reduce inter-machine messaging.  Figure~\ref{fig:gre} shows that an extra \textit{scatter} vertex, $C^\prime$, is added to create the internal Agent-Graph model. The $C^\prime$ vertex acts as a Receiver-side Scatter depicted in Figure~\ref{fig:rec_scat}.  This is useful because the new $C^\prime$ vertex allows $C$ to only send one message across the network, which $C^\prime$ then disperses to vertices $D$, $E$, and $F$.  Combiner vertices are also added to the Agent-Graph in the same way as Server-side Aggregation depicted in Figure~\ref{fig:send_agg}.  The Agent-Graph employed by GRE is similar to vertex-cuts in PowerGraph except that GRE messaging is unidirectional, and active messages are also utilized for parallel graph computation in the Active Pebbles framework \cite{Willcock2011,Edmonds2013}.

\subsubsection{Message Passing Optimizations}
\label{subsubsec:opts}

\begin{figure*}
    \centering
    
    \begin{subfigure}[b]{.3\textwidth}
        \centering
        \raisebox{5mm}{\includegraphics[width=\textwidth]{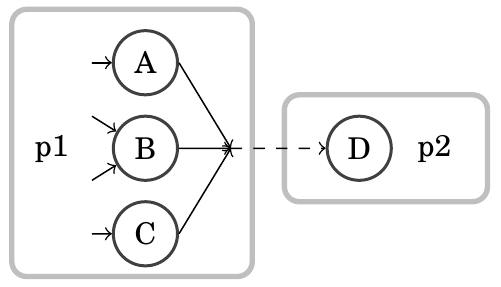}}
        \caption{Sender-side Combiner}
        \label{fig:send_agg}
    \end{subfigure}
    ~
    \begin{subfigure}[b]{.3\textwidth}
        \centering
        \includegraphics[width=.9\textwidth]{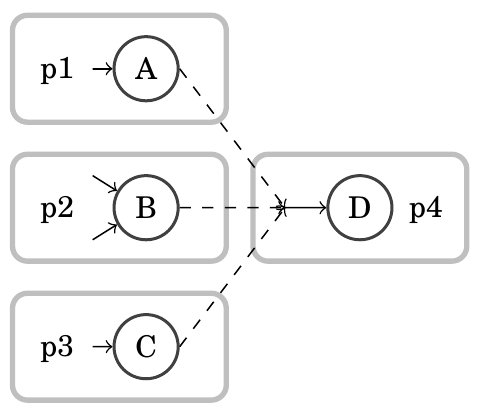}
        \caption{Receiver-side Combiner}
        \label{fig:rec_agg}
    \end{subfigure}
    ~
    \begin{subfigure}[b]{.3\textwidth}
        \centering
        \raisebox{4.5mm}{\includegraphics[width=.9\textwidth]{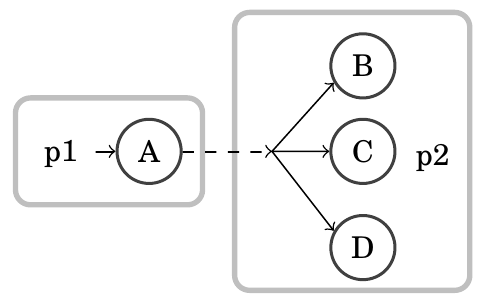}}
        \caption{Receiver-side Scatter}
        \label{fig:rec_scat}
    \end{subfigure}
    
    \caption{Partition-driven optimization strategies for distributed message passing.  The Combiner technique employs both Sender-side and Receiver-side Combiners.}
    \label{fig:opt}
\end{figure*}

Message passing can be costly, especially over a network. Thus several message-reducing strategies have been developed in order to improve performance.  Some strategies are topology-driven and, as such, exploit the graph layout across machines, while other techniques are applied to specific algorithmic behavior.  Three topology-driven optimizations are depicted in Figure~\ref{fig:opt} for messaging between machines $p1$ and $p2$ (or messaging from $p1$, $p2$, and $p3$ to $p4$, for Figure~\ref{fig:rec_agg}). 

The Combiner, inspired by the MapReduce function of the same name \cite{Dean2008}, is a message passing optimization originally used by Pregel \cite{Malewicz2010}.  Presuming the commutative and associative properties of a vertex function, a Combiner executes on a worker process and combines many messages destined for the same vertex into a single message.  For example, if a vertex function computes the sum of all incoming messages, then a Combiner would detect all messages destined for a vertex $v$, compute the sum of the messages, then send the new sum to $v$.  A Combiner can especially reduce network traffic when $v$ is remote, shown as  {\em sender-side aggregation} (Figure~\ref{fig:send_agg}).  When $v$ is local, a combiner can still reduce memory overhead by aggregating messages before placement into the incoming message queue, shown as {\em receiver-side aggregation} (Figure~\ref{fig:rec_agg}). For the single-source shortest path algorithm, a combiner implementation resulted in a four-fold reduction in network traffic \cite{Malewicz2010}.

A related technique is the {\em receiver-side scatter}.  For instances where the same message is sent to multiple vertices on the same remote machine, network traffic can be reduced by sending only one message and then having the destination worker distribute multiple copies, depicted in Figure~\ref{fig:rec_scat}.  The strategy has been employed in multiple frameworks, including the \textit{Large Adjacency List Partitioning} in GPS \cite{Salihoglu2013}, IBM's X-Pregel \cite{Bao2013}, as the \textit{fetch-once} behavior in LFGraph \cite{Hoque2013}, and through \textit{scatter} nodes of the Agent-Graph in GRE \cite{Yan2013}.  The technique reduces network traffic by increasing memory and processing overhead, as worker-nodes must store the out-going adjacency lists of other workers.  With this in mind, GPS maintains a threshold where receiver-side scatter would only be applied for vertices above a certain degree.  Experiments showed that as the threshold is lowered, network traffic at first decreases then plateaus, while runtime decreases but then increases, demonstrating the existence of an optimal vertex-degree threshold.  In X-Pregel, a ten-fold reduction in network traffic from Receiver-side Scatter resulted in a 1.5 times speedup \cite{Bao2013}. Clearly, the receiver-side scatter strategy can be effective, but unlike the combiner is not guaranteed to improve performance.

The three partition-driven optimizations in Figure~\ref{fig:opt} are related to the messaging structure of a framework, and not specific to algorithm behavior, albeit some assumptions are made regarding message computation.  Computation for the combiner must be commutative and associative because order cannot be guaranteed, while messages for the receiver-side scatter must be identical, and independent of the adjacency list.  Still, the techniques are oriented around partition-level messaging and apply to the worker process, only requiring certain operational properties in order to work.  The Message-Online-Computing model proposed in \cite{Zhou2014}, which improves memory usage by processing messages in the queue as they are delivered, also requires operations be commutative.

Conversely, algorithm-specific message optimizations have also been developed that restructure vertex messaging patterns for certain algorithmic behaviors \cite{Salihoglu2014,Quick2012}.  For algorithms that combine vertices into a supervertex, like Boruvka's Minimum Spanning Tree \cite{Chung1996}, the Storing Edges at Subvertices (SEAS) optimization implements a subroutine where each vertex tracks its parent supervertex instead of sending adjacency lists \cite{Salihoglu2014}.  For algorithms where vertices remove edges, like in the 1/2-approximation for maximum weight matching \cite{Preis1999}, the Edge Cleaning on Demand (ECOD) optimization only deletes stale edges when, counter-intuitively, activity is requested for the stale edge \cite{Salihoglu2014}.  To avoid slow convergence, ECOD is only employed above a certain threshold, {\em e.g.}, when more than 1\% of all vertices are active.  Both SEAS and ECOD exploit a trade-off between sending messages proportional to the number of vertices or proportional to the number of edges.  Other strategies for reducing communication, based on {\em aggregate computation}, are discussed in Section~\ref{subsec:scope_opt}. 

\subsection{Execution Model}
\label{subsec:comp_models}

The model of execution for vertex-centric programs describes the implementation of the vertex function, and how data moves during computation.  

\subsubsection{Vertex Program Implementation}

Vertex functions have been implemented as 1, 2, or 3 phase-models.  Vertex functions have also been implemented as edge-centric functions.  While the model choice does not typically impact the accuracy of the final result, combining certain implementations with other TLAV components can yield improved system performance for certain graph characteristics.

\paragraph{One Phase}  

The vertex programming abstraction implemented as a single function is well-characterized by the Pregel framework \cite{Malewicz2010}.  The single compute function of a vertex object follows the general sequence of accessing input data, computing a new vertex value, and distributing the update.  In a typical Pregel program, the input data is accessed by iterating through the input message queue (messages that may have utilized a combiner), applying an update function based on received data, and then sending the new value through messages addressed by iterating over outgoing edges.  Details based on other design decisions may vary, {\em e.g.}, input and output data may be distributed through incident edges, or neighboring vertex data may be directly accessible, but in one-phase models the general sequence of vertex execution is performed within a single, programed function.  The \texttt{Vertex.Compute()} function is implemented in several TLAV frameworks in addition to Pregel, including its open-source implementations \cite{Avery2011,Seo2010} and several related variants \cite{Salihoglu2013,Bao2013,Redekopp2013}.  The One-phase function implementation is conceptually straight-forward, but other frameworks provide opportunities for improvement by dividing up the computation.

\paragraph{Two Phase} 

A two-phase vertex-oriented programming model breaks up vertex programming into two functions, most commonly referred to as the Scatter-Gather model.  In Scatter-Gather, the \textit{scatter} phase distributes a vertex value to neighbors, and the \textit{gather} phase collects the inputs and applies the vertex update.  While most single-phase frameworks {\em e.g.}, Pregel, can be converted into two phases, the Scatter-Gather model was first explicitly put forward in the Signal/Collect framework \cite{Stutz2010}. The two phase model is also presented as Scatter-Gather in \cite{Roy2013}, and is presented as the Iterative Vertex-Centric (IVEC) programming model in \cite{Yoneki2013}.  The Scatter-Gather programming model commonly occurs in TLAV systems where data is read/written to/from edges.  

Ligra and Polymer are frameworks implemented for single-machines (see Section~\ref{sec:arch}) that both implement a two-phase model.  The user provides two functions, one function that executes across each vertex in the active subset and another function that executes all outgoing edges in the subset.  The frameworks adopt a vertex-subset-centric programming model, which is similar to vertex-centric, but the framework retains a centralized view of the graph, where the whole graph is within the scope of computation, which is possible because the entire graph resides on a single machine in this case.  The two phase model is executed within a program processing the whole graph.

A related two-phase programming model for message passing called Scatter-Combine is implemented in the GRE framework \cite{Yan2013}.  This model utilizes active messages, which are messages that include both data as well as the operator to be executed on the data \cite{VonEicken1992}.  In the first phase of the model, messages are both sent (Scattered) and the operators in the messages are executed (Combined) at the destination vertex.  In the second phase, the combined result is used to update the vertex value.  The Scatter-Combine model incorporates two phases differently than Scatter-Gather.  Instead of the two phase Scatter-Gather model of (i) Gather-Apply, and (ii) Scatter, the Scatter-Combine model uses active messages to institute (i) Scatter-Gather, and then (ii) Apply.  The GRE framework combines  Scatter-Combine with a novel representation of the underlying data graph, called the Agent-Graph, described above, to reduce communication and improve scalability for processing graphs with scale-free degree distributions.

\paragraph{Three Phase}  
\label{subsubsec:gas}
A three-phase programming model is introduced in PowerGraph as the Gather-Apply-Scatter (GAS) model \cite{Gonzalez2012}.  The \textit{Gather} phase performs a generic summation over all input vertices and/or edges, like a commutative associative combiner.  The result is used in the \textit{Apply} phase, which updates the central vertex value.  The \textit{Scatter} phase distributes the update by writing the value to the output edges.  PowerGraph incorporates the GAS model with vertex-cut partitioning (see Section~\ref{subsubsec:vert_cuts}) to improve processing of power-law graphs.

\paragraph{Edge-Centric} 
The X-Stream framework provides an \textit{edge-centric} two phase Scatter-Gather programming model \cite{Roy2013}, as opposed to a vertex-centric programming model.  The model is edge-centric because the framework iterates over edges of the graph instead of vertices.  However, the framework may still be considered TLAV because the two phase program operates on source and target vertices, adopting a similar local scope.  X-Stream leverages streaming edge data instead of random access for efficient large scale graph processing on a single machine, and is discussed in Section~\ref{sec:arch} in further detail.

\subsubsection{Push vs. Pull}
\label{subsubsec:pushpull}

The flow of information for vertex-programs can be characterized as data being \textit{pushed} or \textit{pulled} \cite{Nguyen2013,Hant2014,Cheng2012}.  In \textit{push} mode, information flows from the active vertex performing the update outward to neighboring vertices, as in Pregel-like message-passing.  In \textit{pull} mode, information flows from neighboring vertices inward to the active vertex, as in GraphLab-like shared memory, when an active vertex reads neighbor's data.  Few TLAV frameworks explicitly adopt a push or pull mode.  Instead, the information flow arises from other design decisions.  Still, analyzing a system as push or pull allows one to reason about other system properties.  For example, asynchronous execution is supported by both modes, but sender-side combining is only possible in push mode \cite{Cheng2012}.

Push and pull modes are more commonly associated with databases and transactional processing, though have been more explicitly incorporated in broader graph engines and temporal frameworks  (see Section~\ref{sec:related} for related work).  The Galois framework, with a flexible computation model enabling the implementation of a vertex-centric interface, allows users to choose push or pull mode \cite{Kulkarni2007,Nguyen2013}, as does Kineograph \cite{Cheng2012}.  Chronos experiments with how push and pull modes impact caching \cite{Hant2014}.

Ligra is a single-machine graph processing framework that dynamically switches between push and pull-based operators based on a threshold.  The framework is in part inspired by a recently developed shared-memory breadth-first search algorithm that achieves remarkable performance by switching between push and pull modes of exploration \cite{Beamer2013}.  This algorithm, Ligra, and PowerSwitch from Section~\ref{subsubsec:hybrid} exemplify how performance can be improved by dynamically adapting the processing technique to properties of the graph.

The delta-caching optimization, which is introduced in PowerGraph \cite{Gonzalez2012}, which reduces the pulling of redundant data by tracking value changes.  In a three phase model, an accumulator value is the result of gather step.  With delta-caching, a cached copy of the accumulator for each vertex is stored by the worker, requiring additional storage.  If, for a given update, the change in the accumulator is minimal, then neighboring vertices aren't activated, and any change can be applied to the cached copy stored by worker.  A neighboring vertex can then use the cached copy during an update.  For delta-caching to be available, the apply function must be commutative, associative, and have an inverse function.  Delta-caching reduces redundant pulling by not activating neighboring vertices for small changes, and resulted in a 45\% decrease in runtime for computing PageRank on the Twitter graph \cite{Gonzalez2012}.

\subsection{Partitioning}
\label{subsec:partitioning}

Large-scale graphs must be divided into parts to be placed in distributed memory.  Good partitions often lead to improved performance \cite{Salihoglu2013}, but expensive strategies can end up dominating processing time, leading many implementations to incorporate simple strategies, such as random placement \cite{Jain2013}.  Effective partitioning evenly distributed the vertices for balanced workload, while minimizing inter-partition edges to avoid costly network traffic, a problem formally known as {\em k-way graph partitioning} that is NP-complete with no fixed-factor approximation \cite{Andreev2006,Meyerhenke2014}.  

Leading work in graph partitioning can be broadly characterized as (1) rigorous but impractical mathematical strategies, or (2) pragmatic heuristics used in practice \cite{Tsourakakis2014}. Practical strategies, such as those employed in the suite of algorithms known as METIS \cite{Karypis1995a}, often employ a three-phase multi-level partitioning approach \cite{Abou-Rjeili2006}.  Partition size is often allowed to deviate in the form of a ``slackness" parameter in exchange for better cuts \cite{Karypis1996}.  

Graph partitioning with METIS partitioning software is often considered the \textit{de facto} standard for near-optimal partitioning in TLAV frameworks \cite{Stanton2012}.  Despite a lengthy preprocessing time, METIS-algorithms significantly reduce total communication and improve overall runtime for TLAV processing on smaller graphs \cite{Salihoglu2013}.  However, for graphs of even medium-size, the high computational cost and necessary random access the entire graph renders METIS and related heuristics impractical.  Alternatives for large-scale graph partitioning include distributed heuristics presented in Section~\ref{subsubsec:dist_heur}, streaming algorithms in Section~\ref{subsubsec:streaming}, vertex cuts in Section~\ref{subsubsec:vert_cuts}, and dynamic repartitioning in Section~\ref{subsubsec:dyn_repart}.

\subsubsection{Distributed Heuristics}
\label{subsubsec:dist_heur}
Distributed heuristics are decentralized methods, requiring little or no centralized coordination.  Distributed partitioning is related to distributed community detection in networks \cite{Gehweiler2010,Ramaswamy2005}, the two main differences being: 1) communities can overlap whereas partitions cannot, and 2) partitioning requires \textit{a priori} specification of the number of partitions, whereas community detection typically does not.  Much distributed partitioning work has been inspired by distributed community detection, namely label propagation \cite{Raghavan2007}.

Label propagation occurs at the vertex level, where each vertex adopts the label of the plurality of its neighbors.  Though the process is decentralized, label propagation for partitioning necessitates a varying amount of centralized coordination in order to maintain balanced partitions and prevent ``densification": a cascading phenomenon where one label becomes the overwhelming preference \cite{Raghavan2007}. The densification problem is addressed in \cite{Vaquero2013} wherein a simple capacity constraint is enforced that is equal to the available capacity of the local worker divided by the number of non-local workers.  In \cite{Ugander2013}, balanced vertex distribution is maintained by constraining label propagation and solving a linear programming optimization problem that maximizes a relocation utility function. In \cite{Rahimian2013}, vertices swap labels, either with a neighbor or possibly a random node, and simulated annealing is employed to escape local optima.  The cost of centralized coordination incurred by these methods is much less than the cost of random vertex access on a distributed architecture, as with ParMETIS.  

More advanced label propagation schemes for partitioning are presented in \cite{Wang2014} and \cite{Madduri}.  In \cite{Wang2014}, label propagation is used as the coarsening phase of a multi-level partitioning scheme, which processes the partitioning in blocks to accommodate multi-level partitioning for large-scale graphs.  In \cite{Madduri}, several stages of label propagation are utilized to satisfy multiple partitioning objectives under multiple constraints.  \cite{Zeng2012} use a parallel multi-level partitioning algorithm for k-way balanced graphs that operates in two phases: an aggregate phase that uses weighted label propagation, and then a partition phase that performs the stepwise minimizing RatioCut method.

\subsubsection{Streaming}
\label{subsubsec:streaming}

Streaming partitioning is a form of online processing that partitions a graph in a single-pass.  For TLAV frameworks, streaming partitioning is especially efficient since the partitioning can be performed by the graph loader, which loads the graph from disk onto the cluster.  The accepted streaming model assumes a single, centralized graph loader that reads data serially from disk and chooses where to place the data amongst available workers \cite{Stanton2012,Tsourakakis2014}.  Centralized streaming heuristics can be adapted to run in parallel \cite{Stanton2012}, however, depending on the heuristic, concurrency between the parallel partitioners would likely be required \cite{Nishimura2013}.  One of the first online heuristics was presented by Kernighan and Lin and is used as a subroutine in METIS \cite{Kernighan1970}.  GraphBuilder \cite{Jain2013} is a a similar library that, in addition to partitioning, supports an extensive variety of graph loading-related processing tasks.  
A streaming partitioner on a graph loader reads data serially from disk, receiving one vertex at a time along with its neighboring vertices.  In a single look at the vertex the streaming partitioner must decide the final placement for the vertex on a worker partition, but the streaming partitioner has access to the entire subgraph of already placed vertices.  In a variant of the streaming model, the partitioner has an available storage buffer with a capacity equal to that of a worker partition, so the partitioner may temporarily store a vertex and decide the partitioning later \cite{Stanton2012}, however this buffer is not utilized by the top performing streaming partitioners.  For most heuristics, the placement of later vertices is dependent on placement of earlier vertices, so the presentation order of vertices can impact the partitioning. Thus, an adverse ordering can drastically subvert partitioning efforts, however, experiments demonstrate that performance remains relatively consistent for breadth-first, depth-first, and random orderings of a graph \cite{Stanton2012,Tsourakakis2014}.

Two top-performing streaming partitioning algorithms are greedy heuristics.  The first is linear deterministic greedy (LDG), a heuristic that assigns a vertex to the partition with which it shares the most edges while weighted by a penalty function linearly associated with a partition's remaining capacity.  The LDG heuristic is presented in \cite{Stanton2012}, where 16 streaming partitioning heuristics are evaluated across 21 different data sets.  The use of a buffer in addition to the LDF heuristic has been adapted for streaming partitioning of massive Resource Description Framework (RDF) data \cite{Wang2013a}. Another variant uses {\em unweighted} deterministic greedy instead of linear deterministic greed (LDG), to perform greedy selection based on neighbors without any penalty function; this unweighted variant has been employed for distributed matrix factorization \cite{Ahmed2013}.  Further analysis of LDG-related heuristics on random graphs, as well as lower bound proofs for random and adversarial stream ordering, is presented in \cite{Stanton2014}.  

Another top-performing streaming partitioner is FENNEL \cite{Tsourakakis2014}, which is inspired by a generalization of optimal quasi-cliques \cite{Tsourakakis2013}.  FENNEL achieves high quality partitions that are in some instances comparable with near-optimal METIS partitions.  Both FENNEL and LDG have been adapted to the restreaming graph partitioning model, where a streaming partitioner is provided access to previous stream results \cite{Nishimura2013}.  Restreaming graph partitioning is motivated by environments such as online services where the same, or slightly modified, graph is repeatedly streamed with regularity.  Despite adhering to the same linear memory bounds as a single-pass partitioning, the presented restreaming algorithms not only provide results comparable to METIS, but are also capable of partitioning in the presence of multiple constraints and in parallel without inter-stream communication.

\subsubsection{Vertex Cuts}
\label{subsubsec:vert_cuts}
A vertex-cut, depicted in Figure~\ref{fig:powergraph}, is equivalent to partitioning a graph by edges instead of vertices.  Partitioning by edges results in each edge being assigned to one machine, while vertices are capable of spanning multiple machines.  Only changes to values of cut vertices are passed over the network, not changes to edges. Vertex-cuts are implemented by TLAV frameworks in response to the challenges of finding well-balanced edge cuts in power-law graphs \cite{Abou-Rjeili2006,Leskovec2009}.  Complex network theory suggests power-law graphs have good vertex cuts in the form of nodes with high degree \cite{Albert2000}. A rigorous review of vertex separators is presented in \cite{Feige2008}.

PowerGraph combines vertex-cuts with the three-phase GAS model (Section~\ref{subsubsec:gas}) for efficient communication and balanced computation \cite{Gonzalez2012}.  For vertices that are cut and span multiple machines, one copy is randomly designated the master, and remaining copies are mirrors.  During an update all vertices first execute a gather, where all incoming edge values are combined with a commutative associative sum operation.  Then the mirrors transmit the sum value over the network to the master, which executes the apply function to produce the updated vertex value. The master then sends the result back over the network to the mirrors.  Finally, each vertex completes the update by scattering the result along its outgoing edges.  For each update, network traffic is proportional to the number of mirrors, therefore, breaking up high-degree vertices reduces network communication and helps to balance computation. 

Since its initial implementation in PowerGraph, the vertex-cut approach has been adopted by several other TLAV frameworks.  GraphX is a vertex programming abstraction for the Spark processing framework \cite{Gonzalez2014,Zaharia2010} where the adoption of vertex-cuts demonstrated an 8-fold decrease in the platform's communication cost. GraphBuilder \cite{Jain2013}, an open-source graph loader, supports vertex-cuts and implements grid and torus-based vertex-cut strategies that were later included in PowerGraph.  PowerLyra \cite{Chen2013b} is a modification to PowerGraph that hybridizes partitioning where vertices with a degree above a user-defined threshold are cut, while vertices below the threshold are partitioned using an adaptation of the FENNEL streaming algorithm \cite{Tsourakakis2014}. PowerLyra also incorporates unidirectional locality similar to GRE framework (see Section~\ref{subsubsec:acivemsg}).  BiGraph is a framework developed on PowerGraph that implements partitioning algorithms for large-scale bipartite graphs \cite{Chen2014bi}.  LightGraph \cite{Zhao2014} is a framework that optimizes vertex-cut partitions by using edge-direction-aware partitioning, and by not sending updates to mirrors with only in-edges.

Several edge partitioning analyses and algorithms have recently been developed.  A thorough analysis comparing expected costs of vertex partitioning and edge partitioning is presented in \cite{Bourse2014}. In this study, edge partitioning is empirically demonstrated to outperform vertex partitioning, and a streaming least marginal cost greedy heuristic is introduced that outperforms the greedy heuristic from PowerGraph.  

Centralized hypergraph partitioning, including edge partitioning, is NP-hard, and several exact algorithms have been developed \cite{Biha2011,Kim2012,Hager2014,Sevim}. However, because of their complexity, such algorithms are too computationally expensive and not practical for large-scale graphs.  Centralized heuristics have been shown to be equally impractical \cite{Benlic2013}.  A large-scale vertex-cut approach for bipartite graphs based on hypergraph partitioning is presented in \cite{Miao2013} as part of a vertex-centric program for computing the alternating direction of multipliers optimization technique.  A distributed edge partitioner was developed in \cite{Rahimian2014} that creates balanced partitions while reducing the vertex cut, based on the vertex partitioner in \cite{Rahimian2013}.  Good workload balance for skewed degree distributions can also be achieved with degree-based hashing \cite{Xie2014}.  Finally, as part of a non-vertex-centric BSP graph processing framework, a distributed vertex-cut partitioner is presented in \cite{Guerrieri2014} that uses a market-based model where partitions use allocated funds to buy an edge.

\subsubsection{Dynamic Repartitioning}
\label{subsubsec:dyn_repart}

While an effective partitioning equally distributes vertices among the partitions, for TLAV frameworks, the number of active vertices performing updates on a given superstep can vary drastically over the course of computation, which creates processing imbalances and increases run time.  Dynamic repartitioning was developed to maintain balance during processing by migrating vertices between workers as necessary.

Reasons for changing active vertex sets include topological mutations to the graph and algorithmic execution properties.  Topological mutations may occur if the framework supports dynamic or temporal graphs (see Related Work in Section~\ref{sec:related}).  Topology may also change due to the algorithm, such as graph coarsening \cite{Wang2014}.  

With a static topology, the execution pattern of the algorithm can also change the active vertex set.  While vertex algorithms such as synchronous PageRank execute on every vertex for every superstep, other algorithms introduce dynamism.  \cite{Shang2013} classifies 9 vertex algorithms as either (i) always active, (ii) traversal, or (iii) multi-phase, where the active vertex set of the latter two classifications can vary widely and unpredictably, depending on the graph.  For dynamic repartitioning to prove beneficial, the associated overhead must be less than the additional costs stemming from processing imbalance.

According to \cite{Salihoglu2013}, a dynamic repartitioning strategy must directly address (i) how to select vertices to reassign, (ii) how and when to move the assigned vertices, and (iii) how to locate the reassigned vertices.  Other properties of a strategy include whether coordination is centralized or decentralized, and how the strategy combats ``densification" and enforces vertex balance.  Densification is akin to the rich-get-richer phenomenon, and can occur in greedy or decentralized protocols for partitioning/clustering, where one partition becomes over-populated as the repeated destination for migrated vertices \cite{Vaquero2013a}.  In response, protocols often implement constraints that prevent a partition from exceeding a certain capacity.  The XPregel framework, for example, only permits the worker with the most vertices and edges to migrate vertices \cite{Bao2013}.  

Table~\ref{table:dyn_repart} presents 6 TLAV frameworks that support dynamic repartitioning: GPS \cite{Salihoglu2013}, Mizan \cite{Khayyat2013}, XPregel \cite{Bao2013}, xDGP \cite{Vaquero2013a}, LogGP \cite{Xu2013}, and the Catch the Wind prototype \cite{Shang2013}.  The table includes what active vertex set imbalances are targeted by the frameworks, what metrics are used to identify vertices for reassignment, how reassigned vertices are located after migration, how densification is avoided, and whether the protocol is centralized or decentralized.  

Among the 6 frameworks that implement dynamic repartitioning, all are synchronous, and repartitioning occurs at the end of a superstep, separate from the updates.  When a vertex is selected for migration, the worker must send all associated data to the new worker, including the vertex ID, the adjacency list, and the incoming messages to be processed in the next superstep.  To avoid sending all incoming messages over the network, many dynamic repartitioning frameworks implement a form of delayed migration, where the new worker is recognized as the owner of the migrated vertex, but the vertex value remains on the old worker for an extra iteration in order to compute an update.  With delayed migration, the incoming message queue doesn't need to be migrated, but the new worker still receives new incoming messages \cite{Khayyat2013,Salihoglu2013}.

\begin{table*}[t]
\centering
\small{
\begin{tabular} {l | l l l l l}
Framework & \begin{tabular}{@{}l@{}}Cause of \\ Imbalance\end{tabular} & \begin{tabular}{@{}l@{}}Reassignment \\ Metric\end{tabular} & \begin{tabular}{@{}l@{}}How to Locate \\ Migrated Verts\end{tabular} & \begin{tabular}{@{}l@{}}Densification \\ Avoidance\end{tabular} & Coordination  \\ \hline
GPS  & Algorithm & Sent Msgs & \begin{tabular}{@{}l@{}}Broadcast \\ Vert ID\end{tabular}  & Swap Min-Set  & Decentralized \\
Mizan & Algorithm & \begin{tabular}{@{}l@{}}Sent/Recv Msgs \\ and Run Time \end{tabular}& \begin{tabular}{@{}l@{}}Distributed \\ Hash Table\end{tabular}  & Metric-based Swap & Decentralized\\
XPregel & Algorithm & Sent/Recv Msgs & \begin{tabular}{@{}l@{}}Broadcast \\ Worker ID\end{tabular}  & \begin{tabular}{@{}l@{}}Repartition \\ Largest Worker\end{tabular}& Centralized  \\
xDGP & Topology & \begin{tabular}{@{}l@{}}Labels of \\ Neighbors\end{tabular} & \begin{tabular}{@{}l@{}}Broadcast \\ Worker ID\end{tabular} & \begin{tabular}{@{}l@{}}Fraction of \\ Capacity\end{tabular} & Decentralized  \\
LogGP & Both & Runtime & Lookup Table & \begin{tabular}{@{}l@{}}Repartition Longest- \\ Running Workers\end{tabular} & Centralized \\
\begin{tabular}{@{}l@{}}Catch \\ the Wind\end{tabular} & Algorithm & Sent/Recv Msgs & Lookup Table & Quota & Decentralized\\
\end{tabular}
}
\caption{Feature summary for TLAV frameworks that implement dynamic repartitioning.  Active vertex set imbalance may arise from topology changes, algorithm execution, or both.  A repartitioning strategy includes how to select vertices for reassignment, and how reassigned vertices are later located.  Strategies should also avoid densification, and can be centrally or decentrally implemented.  All implemented frameworks are synchronous.}
\label{table:dyn_repart}
\end{table*}

Though fundamentally sound, many experiments demonstrate that dynamic repartitioning is often not worth the high overhead.  Results in \cite{Bao2013} show that while network I/O is significantly reduced over time, overall runtime shows minor improvements.  Independent tests of GPS show dynamic repartitioning to be detrimental for all cases in \cite{Lu2014}, and similar results are observed for GPS and Mizan in \cite{Han2014}.  However, one major shortcoming in these evaluations is the use of the PageRank algorithm for experimentation.  Dynamic repartitioning is most effective for dynamic active vertex sets, but with PageRank vertices are always active, so dynamic repartitioning performs predictably poorly.  Asynchronous dynamic repartitioning protocols have yet to be explored for TLAV frameworks, but the added complexity and overhead for asynchrony demonstrated in Section~\ref{subsec:execution_policy} suggest that such an implementation is not practical.

\section{Implementation}
\label{sec:implementation}

This section overviews implementation details of TLAV frameworks relating to the distributed environment. These details include system architecture and fault tolerance.  Additionally, TLAV frameworks that employ novel techniques to process large-scale graphs on single machines are surveyed.

\subsection{System Architecture}

TLAV frameworks generally always employ the master-slave architecture.  A master node initializes the slave workers, monitors execution, and manages coordination (and synchronization if invoked) amongst the workers.  Generally, the master is responsible for graph loading and partitioning, but with a network filesystem available, the loading and partitioning can be performed in parallel \cite{Salihoglu2013}.  The master also stores global values, such as aggregators \cite{Malewicz2010}.  The workers each execute a copy of the program on the local partitions and inform the master of runtime status.

One notable exception to the general master-slave architecture is XPregel \cite{Bao2013}, implemented in X10 \cite{Charles2005}.  X10 implements an Asynchronous Partitioned Global Address Space (APGAS), which is a shared address space but with a local structure that enables highly productive distributed and parallel programming.  With APGAS, the number of local ``places" is provided at runtime, which the programmer may utilize as necessary.  XPregel does implement master-slave, but in X10, the master is actually just place 0, sans hierarchy, and opens the door for alternative architectures, like recursive structures.

\subsection{Multi-Core Support}

For multi-core machines, many BSP-based frameworks including Pregel \cite{Malewicz2010} simply assign a partition to a given core, but frameworks can better utilize computational resources through multi-threading.  XPregel \cite{Bao2013} supports multi-threading by dividing a partition into a user-defined number of subpartitions, assigning one thread to each subpartition.  GraphLab \cite{Low2012} implements multi-threading and avoids deadlocks through scheduler restrictions.  GPS \cite{Salihoglu2013} implements 3 types of threads: a thread for vertex computation, a thread for communication, and a thread for parsing.  Cyclops \cite{Chen2014} implements a hierarchical BSP model \cite{Cha2001hbsp} with a split design to parallelize computation and messaging while exploiting locality and avoiding synchronization contention.  Cyclops demonstrates that multi-threading can improve runtime relative to single-threaded execution for the same framework, at the expense of added complexity.

\subsection{Fault Tolerance}
\label{subsec:fault-tol}

Distributed systems must often account for the potential failure of one or more nodes over the course of computation.  When a node fails, a replacement node may become available, but all data and computation performed on the failed node is lost.

Checkpointing is a common fault tolerance implementation, where an immutable copy of the data is written to persistent storage, such as a network filesystem.  Pregel implements synchronous checkpointing, where the graph is copied in between supersteps \cite{Malewicz2010}.  When a failure occurs, the system rolls back to the most recently saved point, all partitions are reloaded, and the entire system resumes processing from the checkpoint.  The partition of the failed node is reloaded to a new replacement node.  If messaging information is also logged, then resources can be saved by only reloading and recomputing data  on the replacement node.  GraphLab \cite{Low2012} implements asynchronous vertex checkpointing, based on Chandy-Lamport \cite{Chandy1985} snapshots, which need not halt the entire program and can result in slightly faster overall execution than synchronous checkpointing, minding certain program constraints.

GraphX is a graph processing library for Apache Spark, which is developed based on the Resilient Distributed Dataset (RDD) abstraction \cite{Gonzalez2014}.  RDDs are immutable, partitioned collections created through data-parallel operators, like map or reduce.  RDDs are either stored externally, or generated in-memory from operations on other RDDs.  Spark maintains the lineage of operations on an RDD, so upon any node failure the RDD can be automatically recovered. GraphX leverages the RDDs of Spark to create a graph abstraction and Pregel interface.

The Imitator \cite{wang-replication} framework implements fault-tolerance based on vertex replicas, or ghosts/mirrors used in shared memory (see Section!\ref{subsubsec:shared}).  The use of replicas for fault tolerance is founded in the observation that the hash partitioning of many real-world directed graphs results in the replication of over 99\% of vertices \cite{wang-replication}.  By replicating \textit{every} vertex, a full copy of the graph can reside in distributed memory, enabling faster recovery times at the expense of relatively little additional memory consumption and network messaging  \cite{wang-replication}.  The efficiency of Imitator is tied to the effectiveness of the partitioning (see Section~\ref{subsec:partitioning}).  Imitator outperforms checkpointing for large graphs distributed over several nodes, when only one replica per vertex is required.  State-of-the-art partitioning methods like METIS, or a smaller number of partitions (Imitator experiments were run on 50 nodes), would likely lead to increased overhead for Imitator.  Also, the number of replicas is tied to the degree of fault tolerance.  To support the failure of {\em k} machines, then {\em k} replicas are required, increasing overhead for each additional failure supported.

A partition-based checkpoint method for fault tolerance is presented in \cite{shen-partitionbased}.  During execution, a recovery executor node collects run-time statistics, and upon failure, uses heuristics to redistribute the partitions.  Checkpointed partitions of the failed nodes can be reassigned amongst both new and old nodes, parallelizing recovery.  Partitions on healthy nodes can also be reassigned for load balancing. 

\subsection{Single Machine Architectures}
\label{sec:arch}

Like MapReduce, TLAV frameworks are advantageous because they are highly scalable while providing a simple programming interface, abstracting away the lower level details of distributed computing.  However, such environments also stipulate the availability of elaborate infrastructure, cluster management, and performance tuning, which may not be available to all users.

Single machine systems are easier to manage and program, but commodity machines do not have the memory capacity to process large-scale graphs in-memory.  This section overviews single machine TLAV frameworks that employ {\em novel} methods to process large-scale graphs.  The main features of the 4 single machine frameworks in this section are presented in Table~\ref{table:singlemachine_frameworks}.

Processing large-scale graphs on a single machine requires either substantial amounts of memory, or storing part of the graph out-of-memory, in which case performance is dictated by how efficiently the graph can be fetched from storage.  In \cite{Shun2013}, it's argued that high-end servers, offering 100GB to 1TB of memory or more, is enough capacity for many real and synthetic graphs reported in the literature.  Such machines would be capable of storing large graphs and executing relatively simple graph algorithms, though more complex algorithms would likely exhaust resources.  

The recommendation service at Twitter \cite{Gupta2013}, which implements a single machine graph processing system with 144 GB of RAM, finds that in practice one edge occupies roughly five bytes of RAM on average.  Compression techniques are further explored for large memory servers in \cite{shun2015}.  Yet, graphs of scale are not practical on lower-end machines containing around 8 to 16 GB of memory \cite{Kyrola2012}.  Accordingly, single machine frameworks have been developed that implement the vertex-centric programming model and process a graph in parts.  Central to many single machine TLAV frameworks are novel data layouts that efficiently read and write graph data to/from external storage.  One common representation is the compressed sparse row format, which organizes graph data as out-going edge adjacency sets, allowing for the fast look-up of outgoing edge, and has been implemented in many state-of-the-art shared memory graph processors \cite{Pearce2010,Hong2011}, including Galois \cite{Nguyen2013}.

\begin{table*}
\centering
\small{
\begin{tabular} {l | c c l}
Framework & Storage Medium & Data Layout & \\ \hline
GraphChi & Disk/SSD & Parallel Sliding Window & \cite{Kyrola2012}\\
X-Stream & Disk/SSD & Streaming Partitions & \cite{Roy2013}\\
FlashGraph & SSD Array & Semi-External Memory with Page Cache & \cite{Zheng2015} \\
PathGraph  & Disk/SSD & Compressed DFS Traversal Trees & \cite{Yuan} \\
\end{tabular}
}
\caption{Single Machine Frameworks}
\label{table:singlemachine_frameworks}
\end{table*}

\paragraph*{GraphChi} The seminal single machine TLAV framework is GraphChi \cite{Kyrola2012}, which was explicitly developed for large-scale graph processing on a commodity desktop.  GraphChi enables large-scale graph processing by implementing the Parallel Sliding Window (PSW) method, a graph data layout previously utilized for efficient PageRank and sparse-matrix dense-vector multiplication \cite{Chen2002,Bender2010}.  PSW partitions vertices into disjoint sets, associating with each interval a shard containing all of the interval's incoming edges, sorted by source vertex.  Intervals are selected to form balanced shards, and the number of intervals is chosen so any interval can fit completely in memory.  A sliding window is maintained over every interval, so when vertices from one shard are updated from in-edges,  the results can be sequentially written to out-edges found in sorted order in the window on other shards.  GraphChi may not be faster than most distributed frameworks, but often reaches convergence within an order of magnitude of the performance of distributed frameworks \cite{Kyrola2012}, which is reasonable for a desktop with an order of magnitude less RAM.  The GraphChi framework was later extended to a general graph management system for a single machine called GraphChi-DB \cite{Kyrola2014}.

Storage concepts for single machine graph processing are further explored in \cite{Yoneki2013} through two directions.  The first project investigates reducing random accesses in SSDs through prefetching, in a project called RASP that later evolved into PrefEdge \cite{Nilakant2014}.  The second project is X-Stream \cite{Roy2013}, an edge-centric single machine graph processing framework that exploits the trade-off between random memory access and sequential access from streaming data.

\paragraph*{X-Stream} Streaming data from any storage medium provides much greater bandwidth than random access.  Experiments on the X-Stream testbed, for example, demonstrate that streaming data from disk is 500 times faster than random access \cite{Roy2013}. X-Stream combines a novel data layout, where an index is built over a storage-based edge list with an edge-centric Scatter-Gather programming model that includes a shuffle phase. Data is read from, and updates are written to, streaming edge data.  Though the framework is edge-centric, a user-defined update function is executed on the destination vertex of an edge.  X-Stream reports that it can process a 64-billion edge graph on a single machine with a pair of 3TB magnetic disks attached \cite{xstreamweb}.

\paragraph*{FlashGraph} While GraphChi and X-Stream are designed for general external storage, the FlashGraph framework is developed for graphs stored on any fast I/O device, such as an array of SSDs.  FlashGraph is deployed on top of the set-associative file system (SAFS) \cite{Zheng2015}, which includes a scalable lightweight page cache, and implements a custom asynchronous user-task I/O interface that reduces overhead for asynchronous I/O.  FlashGraph employs asynchronous message-passing and vertex-centric programming with the semi-external memory (SEM) model \cite{Pearce2010}, where vertices and algorithmic state reside in RAM, but edges are stored externally. In experiments comparing GraphChi and XStream, FlashGraph outperformed both by orders of magnitude even when the data for GraphChi and XStream was placed into RAM-disk \cite{Zheng2015}.  

\paragraph*{PathGraph} In addition to the path-centric programming model, further discussed in Section~\ref{sec:alt}, PathGraph also implements a path-centric compact storage system that improves compactness and locality \cite{Yuan}.  Because most iterative graph algorithms involve path traversal, PathGraph stores edge traversal trees in depth-first search order.  Both the forward and reverse edge trees are each stored in a chunk storage structure that compresses data structure information including the adjacency set, vertex IDs, and the indexing of the chunk.  The efficient computational model and storage structure of PathGraph resulted in improved graph loading time, lower memory footprint, and faster runtime for certain algorithms when compared to GraphChi and X-Stream.

\section{Alternative Graph Granularity}
\label{sec:alt}

The strengths of the vertex-centric programming model are also its weaknesses.  Whereas vertex programs may be relatively simpler to reason about since only local data is available, the algorithms are less expressive than conventional centralized algorithms.  While TLAV frameworks exhibit better scalability, execution can be slow because of high overhead from synchronization and message traffic that takes magnitudes longer compared to computation.  Several frameworks strive for the best of both worlds by adopting a scope that is greater than a vertex but less than the graph, summarized in Table~\ref{table:alt_frameworks}.

\subsection{Subgraph-centric Frameworks}

Considering the challenges addressed by TLAV frameworks, taking a subgraph-centric approach is sensible.  Conventional graph algorithms require the entire graph in memory, which is not possible with graphs of scale.  A subgraph, though, can be partitioned into a size small enough to fit into memory (considering computation) while the connections between subgraphs would be no more, and likely much less, than the total number of edges.  The system would better utilize processing while retaining scalability.

The subgraph-centric programming model is implemented in varying degrees by several frameworks.  The Giraph++ \cite{Tian2013}, Blogel \cite{Yan2014}, and GoFFish \cite{Simmhan2013} frameworks provide a subgraph-centric interface for progrmaming sequential algorithms.  Both Giraph++ and Blogel provide a subgraph-centric interface in addition to a vertex-centric interface.  The results of the sequential programs can then be shared either through vertex programs on boundary nodes, or in the case of Blogel, results can be shared directly between subgraphs.  GoFFish exclusively offers a subgraph-centric interface, and implements messaging between subgraphs and also from subgraphs to specific vertices, the latter being used for traversal algorithms.  By allowing subgraphs to directly message vertices, any vertex-centric algorithm can be implemented by a subgraph-centric framework, maintaining scalability while enabling significant performance improvement.  Collectively, subgraph-centric frameworks dramatically outperform TLAV frameworks, often by {\em orders of magnitude} in terms of computing time, number of messages, and total supersteps \cite{Tian2013,Yan2014}.  

The GraphHP \cite{Chen} and  P++ \cite{Zhou2014} frameworks do not implement an interface for sequential programs, but do differentiate between inter-partition nodes to improve performance.  In these two frameworks, supersteps are split into two phases: in the first phase messages are exchanged between vertices on partition boundaries, and in the second phase, vertices within a partition repeatedly execute the vertex program to completion, exchanging messages in memory.  This method reduces communication and improves performance, however, iteratively executing intra-worker vertex programs is less efficient than executing a sequential algorithm. Message-passing algorithms are typically more scalable than sequential graph algorithms, but P++ is not distributed, nor is Block-based GRACE \cite{Xie2013a}, an extension of \cite{Wang2013}, although the later demonstrates that executing vertex updates on a subgraph block basis improves locality and cache hits while reducing memory access time, which is a bottleneck for computationally light algorithms like PageRank.

TLAV frameworks illustrate the principal ideas for scalable graph processing, but for the best performance, users may consider subgraph-centric frameworks. Subgraph frameworks leverage principles of TLAV frameworks to execute sequential graph algorithms in a distributed environment.  The Giraph++, Blogel, and GoFFish frameworks reduce the scope of sequential graph algorithms for the subgraph to fit in memory while utilizing vertex or subgraph messaging to maintain scalability.  Together, the vertex-centric and subgraph-centric programming model, compared to sequential graph algorithms, demonstrate how scalability varies inversely with scope.  

\begin{table*}
\centering
\small{
\begin{tabular} {l | c c c c l}
Framework & \begin{tabular}{@{}l@{}}Programming \\ Model\end{tabular} & \begin{tabular}{@{}l@{}}Sequential \\ Algorithms\end{tabular} & \begin{tabular}{@{}l@{}}Vertex \\ Messaging\end{tabular} & Distributed & \\ \hline
Giraph++ & Subgraph & Y & Y & Y & \cite{Tian2013} \\
Blogel & Subgraph & Y & Y & Y & \cite{Yan2014} \\
GoFFish & Subgraph & Y & N & Y & \cite{Simmhan2013} \\
GraphHP & Subgraph & N & Y & Y & \cite{Chen} \\ 
P++ & Subgraph & N & Y & N & \cite{Zhou2014} \\
GRACE (block) & Subgraph & N & Y & N & \cite{Xie2013a} \\ 
PathGraph & Path & N & Y & Y & \cite{Yuan} \\
Ligra & Vertex Subset & Y & Y & N & \cite{Shun2013} \\
Polymer & Vertex Subset & Y & Y & N & \cite{Zhang2015} \\
Galois & User-Defined Set & Y & Y & N & \cite{Nguyen2013} \\
\end{tabular}
}
\caption{Frameworks of Alternative Scope}
\label{table:alt_frameworks}
\end{table*}

\subsection{Other Scopes: Paths and Sets}

While subgraph-centric frameworks illustrate the scope/scalability trade-off, several other frameworks adopt alternative computational scopes that demonstrate additional benefits.

A more specific type of subgraph, a traversal tree, is used for the programming model in PathGraph \cite{Yuan}.  Traversals are a fundamental component of many graph algorithms, including PageRank and Bellman-Ford shortest path.  PathGraph first partitions the graph into paths, with each partition represented as two trees, a forward and reverse edge traversal.  Then, for the path-centric computational model, path-centric scatter and path-centric gather functions are available to the user to define an algorithm that traverse each tree.  The user also defines a vertex update function, which is executed by the path-centric functions during the traversal.  Like block-based GRACE, the path-centric model utilizes locality to improve performance through reduced memory usage and efficient caching.  PathGraph also implements a path-centric storage model that enables the framework to process billion node graphs on a single machine (see Section~\ref{sec:arch}) \cite{Yuan}.

Graph processing frameworks designed for single machines can implement interfaces of unique granularity.  A vertex subset interface is implemented in Ligra \cite{Shun2013}.  Ligra argues that high-end servers provide enough memory for large-scale graphs, and thus implements a vertex-centric programming interface while retaining a global view of the graph.  Inspired by a hybrid breadth-first search (BFS) algorithm \cite{Beamer2013}, Ligra dynamically switches between sparse and dense representations of edge sets depending on the size of the vertex subset, which impacts whether push or pull operations are performed with the vertex subset.  Polymer \cite{Zhang2015} adopts a similar interface as Ligra, but with several NUMA-aware optimizations.  Galois \cite{Kulkarni2007} is a shared memory framework that executes user-defined set operators while exploiting amorphous data parallelism \cite{pingali11}.  Galois can be implement a variety of programming interfaces, including the vertex-centric paradigm \cite{Nguyen2013}.

\subsection{Optimizations}
\label{subsec:scope_opt}

Two optimizations have been introduced in \cite{Salihoglu2014} for TLAV frameworks that improve performance by adopting a scope of the graph other than vertex-centric.  The Finishing Computation Serially (FCS) method is applicable when an algorithm with a shrinking set of active vertices converges slowly near the end of execution \cite{Salihoglu2014}.  The FCS method is triggered when the remaining active graph can fit in the memory of a single machine; in these instances the active portions are sent to the master and completed serially from a global, shared memory perspective of the graph. 

Similarly, the Single Pivot (SP) optimization \cite{Salihoglu2014}, first presented in \cite{Quick2012}, also temporarily adopts a global view.  For algorithms that execute breadth-first search (BFS) across all vertices, {\em e.g.}, the connected components algorithm, instead of executing BFS from every node, which incurs a high messaging cost, SP randomly selects one vertex from the graph and performs BFS just from that vertex.  Since most graphs have one big component, in addition to many small ones, the BFS from a random node can be executed until the big component is found, then BFS from every vertex that's not in the big component can execute BFS to complete the algorithm, resulting in significantly fewer total messages.  This optimization adjusts scope by randomly selecting a single vertex by utilizing a global aggregator \cite{Malewicz2010}, which also adopt a scope beyond vertex.

\section{Related Work}
\label{sec:related}

In this paper, vertex-centric graph processing systems for large-scale graphs are surveyed.  In previous related work, Pregel and GraphLab have been compared \cite{Sakr2013}, and general graph processing systems have been surveyed \cite{Khan2014,Nisar2013}, and 4 TLAV frameworks have been empirically evaluated on 4 algorithms \cite{Han2014}.  A tutorial on TLAV frameworks was recently delivered at an international conference \cite{Ajwani2015}.

TLAV frameworks intersect several subjects, including graph processing, distributed computing, Big Data, and distributed algorithms.  Several graph processing frameworks have been recently developed outside of the vertex-centric programming model.  PEGASUS combines the BSP model with generalized matrix-vector multiplication (GIM-V) \cite{Kang2011}, while TurboGraph introduces the pin-and-slide model to perform GIM-V on a single machine \cite{Han2013}. Combinatorial BLAS \cite{Bulucc2011} and the Parallel Boost Graph Library \cite{Gregor2005} are software libraries for high-performing parallel computation of sequential programs.  Piccolo performs distributed graph computation using distributed tables \cite{power2010}.  

Graph databases, such as Neo4j \cite{Webber2012}, HyperGraphDB \cite{Iordanov2010}, and GBASE \cite{Kang2011}, are decidedly different from TLAV frameworks.  Both treat vertices as first class citizens, and both face related problems like partitioning, but the key distinction is that databases focus on transactional processing while TLAV frameworks focus on batch processing \cite{Chen2012}. Databases offer local or online queries, such as 1-hop neighbors, whereas TLAV systems iteratively process the entire graph offline in batch.  Some more general graph management systems, like  Trinity \cite{Shao2013} and Grace \cite{Prabhakaran2012}, offer suites of features that include both vertex-centric processing and queries.  Sensibly, a graph processing engine may be developed on top of a graph database.  However the two should not be confused, and performance is incomparable.

A closely related Big Data framework is MapReduce \cite{Dean2008,Polato2014}.  MapReduce is a different programming model from TLAV frameworks, but similarly enables large-scale computation and, when implemented, abstracts away the details of distributed programming.  The programming model is effective for many types of computation, but addresses neither iterative processing nor graph processing \cite{Polato2014,Malewicz2010}.  Iterative computation is not natively supported, as the programming model performs only a single pass over the data with no loop awareness.  Moreover, I/O is read/written to/from a distributed filesystem, {\em e.g.,} HDFS, rendering iterative computation inefficient \cite{Polato2014}.  Nonetheless, several frameworks have extended MapReduce to support iterative computation \cite{Ekanayake2010,Bu2012,Zhang2012} but such frameworks are still agnostic to the challenges of graph processing.  Graph computation with MapReduce has been explored \cite{Lin2010}, but is generally acknowledged to be lacking \cite{Cohen2009,Malewicz2010}.  A comparison of MapReduce and BSP is provided in \cite{Kajdanowicz2014}.  Still, some argue that MapReduce should remain the sole ``hammer" for Big Data analytics because of the widespread adoption throughout industry \cite{Lin2013}.

Similarly, in response to TLAV shortcomings, such as poor out-of-core support and lengthy loading times, some frameworks rework pre-existing graph database technologies to provide a vertex-centric interface \cite{Fan2015}.  However, many of these projects lose sight of the main problems addressed by the vertex-centric processing.  TLAV frameworks are ultimately Big Data solutions, designed large graphs to be leveraged against the memory and processing power of several machines, not single machines.  Moreover, TLAV frameworks iteratively process the entire graph, and do not provide graph queries like 1-hop or 2-hop neighbors.  TLAV frameworks are not a universal solution for graph analytics, but rather provide an approach for scalable, iterative graph processing.

Temporal graph processing is beyond the scope of this survey, though a small number of TLAV frameworks have been developed for temporal analysis \cite{Cheng2012,Hant2014}.  These frameworks compute temporal properties offline in batch through graph snapshots, necessitating multiple framework components, including a front-end ingress component, an analytics engine, and a storage component such as a graph database.  Temporal  graph layout optimizations were introduced in Chronos \cite{Hant2014}.  These frameworks illustrate how advanced graph analytics systems utilize the strengths of different graph technologies for different components, {\em e.g.,} graph databases for storage and online queries, and vertex-centric computation for batch analytics.  Dynamic graph algorithms and general analytics systems have also been surveyed \cite{Aggarwal2014,Vaquero2014}.  Dynamic graphs are supported by many frameworks including Pregel, but the topic was omitted from this survey due to widely varying support by the frameworks and broad scope of the topic.

While coined "vertex-centric" relative to conventional graph processing approaches, the algorithms executed by TLAV frameworks are more formally known as distributed algorithms.  Distributed algorithms is a mature field of study \cite{Lynch1996}, and further examples beyond Figure~\ref{fig:minval} may be found within the referenced frameworks.  Some works have explored distributed algorithms within the context of TLAV frameworks \cite{Yan2013}, but researchers and practitioners should be aware that TLAV frameworks execute distributed algorithms \cite{Lynch1996}, which come from a field with a considerable body of work, including theory and analysis.  The theoretical limits of what can be computed with vertex-centric frameworks, specifically with the synchronous, message-passing LOCAL model, has been studied \cite{kuhn2010}.  

This paper surveys and compares the various components of TLAV frameworks, which are a platform for executing vertex-centric algorithms.  Like MapReduce, these frameworks provide an interface for a user-defined function, while abstracting away the lower-level details of cluster computing.  Changing the components of the framework will impact system performance and run-time characteristics, but will generally not impact the design or result of the algorithm \footnote{An exception to this rule is synchronous versus asynchronous execution some algorithms, such as graph coloring Section~\ref{subsec:execution_policy}.}.

\section{Conclusions}
\label{sec:conc}

TLAV frameworks have been designed in response to the challenges of processing large graphs.  Primary challenges include the unstructured nature of graphs, where an edge may span any two vertices, so the entire graph must be randomly accessible for conventional processing.  TLAV frameworks are also developed for ease of use, providing a simple vertex-centric interface while abstracting away the lower level details of cluster computing.  MapReduce similarly enables highly scalable computing, but is ill-suited for iterative graph processing.

By adopting a vertex-centric programming model, the scope of computation is dramatically reduced.  To perform an update, each vertex only needs data from immediate neighbors.  Data residing on a separate machine can be acquired directly between workers, avoiding the bottleneck of central coordination, enabling excellent scalability.  The four pillars of the vertex-centric programming model, (i) timing, (ii) communication, (iii) the execution model, and (iv) partitioning, were presented and surveyed in the context of distributed graph processing frameworks.  However, vertex-centric algorithms, colloquially known as distributed algorithms, have an established history and are still actively researched \cite{Lynch1996,kuhn2010}.

Several related frameworks were explored that similarly adopt a computational scope of the graph at varying granularity.  These frameworks of alternative scope are like a Goldilocks solution to graph processing.  Centralized algorithms with the entire graph in scope require too much memory, vertex-centric algorithms can scale but are less expressive and require many relatively slow messages, whereas subgraph-centric algorithms can utilize the two resources just right.  A significant contribution of TLAV frameworks is exposing how, for graphs, reducing the scope of a program increases scalability.  

Of course, expressing a particular algorithm as subgraph-centric is not trivial.  The future of practical large-scale distributed graph processing may be related to finding algorithms that process a graph as independent subgraphs, such as divide-and-conquer, or algorithms that can process graphs at multiple, or even dynamic, scopes \cite{Wang2014}.  The performance of the subgraph-centric processing is also closely tied to the effectiveness of large-scale graph partitioning, including streaming and distributed partitioning techniques.

TLAV frameworks are a tool for graph processing at scale.  Not all graphs are large enough to necessitate distributed processing, and not all graph problems need the whole graph to be computed iteratively.  Moreover, there is often more than one way to solve a problem, but these frameworks are simple to program, easy to distribute, and are not a bad choice for the right type of problem.  Subgraph-centric frameworks take vertex-centric frameworks a step further for performance.  Datasets will continue to grow dramatically into the new age of Big Data, and the design of processing systems should begin asking if they can scale out infinitely.  TLAV frameworks illustrate how conventional centralized systems will fail in the Big Data ecosystem, and how decentralized platforms must be embraced.

\section{Acknowledgements}
This work was supported in part by the AFOSR Grant \#FA9550-15-1-0003, as well as a Department of Education GAANN Fellowship awarded by the University of Notre Dame Department of Computer Science and Engineering.


\begin{thebibliography}{00}
\bibitem[\protect\citeauthoryear{Lin}{Lin}{2013}]%
        {Lin2013}
{Jimmy Lin}. 2013.
\newblock \showarticletitle{Mapreduce is good enough? if all you have is a
  hammer, throw away everything that's not a nail!}
\newblock {\em Big Data\/} {1}, 1 (2013), 28--37.
\newblock
\end{thebibliography}
\end{document}